\documentclass{iopart}
\usepackage{iopams}  
\usepackage[colorlinks]{hyperref} 
\usepackage{graphicx}
\usepackage[dvips]{color}
\usepackage{amssymb}
\usepackage{iopams}  
\usepackage{mathptm}
\usepackage{wasysym}
\usepackage{anysize,color}
\usepackage{graphicx}

\marginsize{1in}{1in}{1in}{1in}

\newcommand{\be}{\begin{equation}}
\newcommand{\ee}{\end{equation}}
\newcommand{\ket}[1]{|#1\rangle}
\newcommand{\bra}[1]{\langle#1|}

\begin{document}

\title[Simulations of quantum double models]{Simulations of quantum double models}

\author{G.K. Brennen$^{1}$, M. Aguado$^{2}$, and J.I. Cirac$^{2}$. }
\address{$^{1}$Centre for Quantum Information Science and Security, Macquarie University, 2109, NSW Australia\\
$^{2}$Max-Planck-Institut f\"ur Quantenoptik, Hans-Kopfermann-Str. 1, D-85748 Garching Germany}

\begin{abstract}
 We demonstrate how to build a simulation of two dimensional physical
 theories describing topologically ordered systems whose excitations
 are in one to one correspondence with irreducible representations of
 a Hopf algebra, $\mathrm{D}(G)$, the quantum double of a finite group $G$.
 Our simulation uses a digital sequence of operations on a spin
 lattice to prepare a ground ``vacuum'' state and to create, braid
 and fuse anyonic excitations.  The simulation works with or without
 the presence of a background Hamiltonian though only in the latter
 case is the system topologically protected.  We describe a physical
 realization of a simulation of the simplest non-Abelian model,
 $\mathrm{D} ( S_3 )$, using trapped neutral atoms in a two dimensional optical
 lattice and provide a sequence of steps to perform universal
 quantum computation with anyons.  The use of ancillary spin degrees
 of freedom figures prominently in our construction and provides a
 novel technique to prepare and probe these systems.
\end{abstract}
\pacs{03.67.Lx, 03.65.Vf, 37.10.Jk}
\maketitle
\tableofcontents
\section{Introduction}

Broadly speaking, topology deals with properties of spaces invariant
under continuous transformations.  Topology appears in a physical
setting, typically, when a quantity $X$ can be shown to take values in
a discrete set $\{ x_1, \, x_2, \, \ldots \}$ when the parameters of a
configuration space $\mathcal{F}$ vary over a continuous range, in
which case $\mathcal{F}$ splits into sectors labelled by the
discrete values of $X$,
\begin{equation}
 \mathcal{F}
=
 \bigcup_{ x_i \in X } {\mathcal{F}}_i
 \; .
\end{equation}
these values providing a \emph{topological classification} of
configurations.  Two configurations in different
sectors cannot be transformed into one another continuously, in other
words, quantity $X$ is \emph{topologically stable} and is, for
instance, conserved during (smooth) time evolutions both in the
classical and quantum worlds.  Examples range from soliton theory in
hydrodynamics, where soliton charges determine the stability of
solitary waves, to the classification of instantons, the key to chiral
symmetry breakdown in four-dimensional Yang-Mills theory.

\emph{Topological phases of matter} and lattice systems
\cite{wen-order} are highly correlated phases whose order cannot be
described in the local group symmetry and order parameter paradigm.
Such systems exhibit a \emph{topological order} usually associated
with a gapped ground level with degeneracy dependent on the
topological properties (typically, a Betti number) of the underlying
spatial manifold --- and, in two dimensions, with particle-like
excitations with anyonic statistics, and gapless edge modes when
boundaries are present.  Fractional quantum Hall states exhibit such
an order and provide an experimental setting where Abelian anyonic
statistics have been shown to exist, and non-Abelian statistics are
widely expected to appear for certain filling fractions.
Topologically ordered lattice systems have also been theoretically
constructed, and their experimental simulation complements their
analytical and numerical study.

Topological orders have recently attracted considerable interest in
the field of quantum information, due to proposals to use topological
phases of matter and lattice systems as quantum memories, whereby
information is stored in topologically stable quantities, and quantum
computers.  In the most widely explored scenario, gates are performed
on the stored information by creation, braiding and fusion of anyonic
excitations.  While it is not clear whether usual topological codes in
two dimensions remain useful at finite temperatures or in the
 presence of noise \cite{fragility}, models in two dimensions have been
proposed featuring string tensions between anyons
\cite{bombin-nonabelian}, therefore confining anyonic defects --- on
the other hand, a kind of topological memories in four dimensions is
also expected to show resilience to thermal effects.  The anyon
braiding paradigm of topological quantum computation hence deserves to
be further studied and realised in the laboratory.

In this respect, it is instructive to consider, following DiVincenzo's
example, the requirements an experimental setting must fulfil in order
to implement anyonic topological quantum computation based on anyon
braiding (see also \cite{gavin:why}):
\begin{itemize}
\item %
 \textbf{Existence}:
 \begin{itemize}
 \item[\textbf{E0}] %
   Existence of anyons.

   Anyonic statistics is used to act on code quantum states by
   manipulating anyonic objects.  Quantum information may also be
   encoded in static configurations of anyons.


 \item[\textbf{E1}] %
   Controlled initialisation of anyons.

   Anyons may be present from the beginning in the system, or may
   need to be created; manipulation of anyons by transport or fusion
   may be required to bring the system to the desired initial
   configuration (e.g., a product state.)


 \item[\textbf{E2}] %
   Implementation of gates by anyon braiding.

   Braiding non-Abelian anyons results in the application of quantum
   gates.  The statistics should be rich enough to provide a
   universal set of braiding quantum gates, in which case we speak of
   universal anyons.


 \item[\textbf{E3}] %
   Measurement of topological charge.

   This is the read-out part of a computation, and can be achieved in
   a number of ways, e.g., by using interferometry.  Such methods may
   involve the controlled fusion of anyons.
 \end{itemize}
\item %
 \textbf{Experimental feasibility}: These are not independent of the
 existence requirements, since, e.g., the ability to perform gates
 imply that the life of an anyon is long enough to carry out the
 task, but their importance as criteria for practical implementations
 warrants a separate listing.
 \begin{itemize}
 \item[\textbf{S0}] %
   Scalability of the implementation.

   A practical implementation should maximise the encoded information
   while ensuring control of the anyonic population.  This leads to
   spatial efficiency as a design goal.
 \item[\textbf{S1}] %
   Robustness of the implementation.

   For instance, the anyon lifetime should be larger than the gate
   application time, or than computation time for
   information-encoding anyons.


 \end{itemize}
\end{itemize}
The basic tools to perform anyonic TQC can be seen to be the
controlled creation, transport, and fusion of anyons.  The status of
anyonic computation in different experimental settings is as follows:
\begin{itemize}
\item %
 \emph{Optical lattices}: satisfy all of \textbf{E0-E3} as argued in
 this article.
\item %
 \emph{Quantum Hall effect}: The existence of Abelian anyons as
 excitations of fractional quantum Hall systems is an established
 fact.  Non-Abelian excitations have been argued to appear for
 filling fraction $5/2$ and $12/5$ --- in the latter case, braiding
 would be universal.  See recent experimental progress in
 \cite{dolev}.
\item %
 \emph{Josephson junction arrays}: The theory was put forward in
 \cite{doucot}, leading to the experiments reported in
 \cite{gladchenko}.
\item %
 \emph{Photonic systems}: See \cite{lu:photon}, \cite{pachos:photon}
 for recent experiments using entangled states of photons.
\end{itemize}

Simulation of lattice systems using cold atoms and molecules in
optical lattices offers an attractive setting where a high degree of
control over the parameters of the system can be achieved.  Minimal
building blocks of topologically ordered systems have been considered
in \cite{paredes}.  The implementation of Kitaev's honeycomb lattice
model was discussed in \cite{duan} and \cite{micheli}; braiding and
interferometry experiments mediated by optical cavity modes were
considered in \cite{jiang}.  Braiding in the Abelian phase was also
considered in \cite{zhang_opt}, however the degradation of anyonic
 visibility with this approach due to the perturbative treatment of
 the honeycomb Hamiltonian was pointed out in \cite{dusuel}, and a
method to construct Kitaev's toric code from the cluster state was
outlined in \cite{Han:07}.

In this work we develop a practical, universal
method, introduced in \cite{Aguado}, to perform anyonic
interferometry and arbitrary quantum computation tasks based on anyon
braiding in spin lattices, using controlled operators applied by
manipulating an extra species of ancillary particles to construct all
needed primitives: anyon creation, transport and fusion.  We
illustrate the method using the $\mathrm{D} ( S_3 )$
quantum double model, whose excitations allow for universal quantum
computation by braiding \cite{mochon}.  Our construction is based on
the spin lattice model introduced by Kitaev \cite{kitaev_toric} with the 
new feature that we give explicit protocols for manipulating anyonic excitations
 and read out of fusion products.  We also describe a method to construct the relevant topologically
ordered states efficiently in absence of a simulated topological
Hamiltonian; however, while this construction is of interest in
itself, we want to emphasise that the anyon manipulation procedure is
independent and can be used whenever the topological phase is
achieved, e.g., by simulation of the Hamiltonian.

\section{\label{qdouble}%
Spin lattice models for the quantum double of a finite group}

Denote $\mathrm{D}(G)$ the quantum double of a finite group
 $G=\{g_j\}$, which is a quasi-triangular Hopf algebra.  An
algebraic construction of these models was first given by Bais et
al. \cite{Bais:93} and for a brief introduction see \ref{qdouble}. 
 We simulate a spin lattice Hamiltonian $H_{\rm TO}$
which is a sum of quasi-local operators and has localized particle
like excitations are in one-to-one correspondence with
the irreducible representations (irreps) of $\mathrm{D}(G)$.  Consider
a two-complex $\Gamma$, which is a cellulation of a
two dimensional surface with vertex set
$\mathcal{V}=\{v_i\}$, edge set $\mathcal{E}=\{e_j\}$, and face set
$\mathcal{F}=\{f_j\}$.  Particles with $d=|G|$ internal
levels (qudits) are placed on the edges and physical states live in a
Hilbert space $\mathcal{H}=\mathcal{H}(d)^{\otimes |\mathcal{E}|}$
where $\mathcal{H}(d)=\mathbb{C}\ket{0}+\cdots +\mathbb{C}\ket{d-1}$.
Particles on edges that meet at a vertex $v$ all interact via a vertex
operator $A_v$.  Similarly, all particles on edges that are on the
boundary of a face $f$ interact via $B_f$.  We pick an orientation for
each edge with $e=[v_j,v_k]$ denoting an edge with arrow pointing from
vertex $v_j$ to $v_k$.  The choice of edge orientations is not
important as long as a consistent convention is used.  We assume an
orientable complex $\Gamma$ and each face $f$ has an orientation
(say, counterclockwise).  The Hamiltonian is a sum of
constraints chosen such that the ground states of $H_{\rm TO}$ are
invariant under local gauge transformations
\be T_g(v)=\prod_{e_j\in
 [v,*]}L_g(e_j)\prod_{e_j\in [*,v]}R_{g^{-1}}(e_j),
\ee
Here $L_g(e_j),R_g(e_j)\in U(d)$ the $d$ dimensional unitary
group, are the permutation representations of the left and
right action of multiplication by the group element $g\in G$ on the
system particle located at edge $e_j$. For the particle states we make
the identification $\ket{j}\equiv\ket{g_j}$, where by convention
$\ket{0}\equiv\ket{g_0}\equiv\ket{e}$, with $e$ the identity
element. The action of left and right group multiplication on the
basis states is then $L_h\ket{j}=\ket{h g_j}$, $R_h\ket{j}=\ket{g_j
 h}$.

A suitable spin lattice model was provided by Kitaev \cite{kitaev_toric}:
\begin{equation}
H_{\rm TO}=-\sum_v A(v)-\sum_f B(f)
\end{equation}
where 
\begin{equation}
\begin{array}{lll}
A(v)&=&\frac{1}{|G|}\sum_{g\in G}T_g(v), \\
B(f)&=&\sum_{\{ \prod_{e_k\in\partial f}h_k=e\}}\otimes_{e_k\in \partial f}\ket{h_k^{-o_f(e_k)}}_{e_k}\bra{h_k^{-o_f(e_k)}}
\end{array}
\end{equation}
In the definition of $B(f)$, the sum is taken over all products of
group elements $h_k$ acting on a counterclockwise cycle of edges on
the boundary of $f$ such that the accumulated left action is the
identity element $e\in G$ (i.e. $h_{\ell}h_{\ell-1}\ldots h_2h_1=e$
for the counterclockwise cycle starting at edge $e_1$ and ending at
edge $e_{\ell}$).  The function $o_f(e_j)=\pm 1$ according to whether
the orientation of the edge is the same as(opposite to) the face
orientation.  By construction
$[A(v),A(v')]=[B(f),B(f')]=[A(v),B(f)]=0$.  Furthermore, it is
straightforward to verify that since $A(v)$ is a symmetrized gauge
transformation it is a projection as is $B(f)$.  The ground states of
$H_{\rm TO}$ are then manifestly gauge invariant states.  Excited
states are described by violations of the local constraints
$A(v),B(f)$ and are particle-like corresponding to the irreps of
$\mathrm{D}(G)$ labeled by $\Pi^{[\alpha]}_{R(N_{[\alpha]})}$ where $[\alpha]$
denotes a conjugacy class of $G$ which labels the magnetic charge, and
$R(N_{[\alpha]})$ denotes a unitary irrep $R$ of the centralizer of an
element in the conjugacy class $[\alpha]$ which labels the electric
charge.  Note there is an arbitrariness in how one picks the fiducial
element of the conjugacy class.  However, $N_{ghg^{-1}}=gN_hg^{-1}$ so
that the centralizers for the elements in a given conjugacy class are
isomorphic (they are equal up to a gauge transformation), and we can
index them just by the conjugacy class.

\begin{figure}
\begin{center}
\includegraphics[scale=0.7]{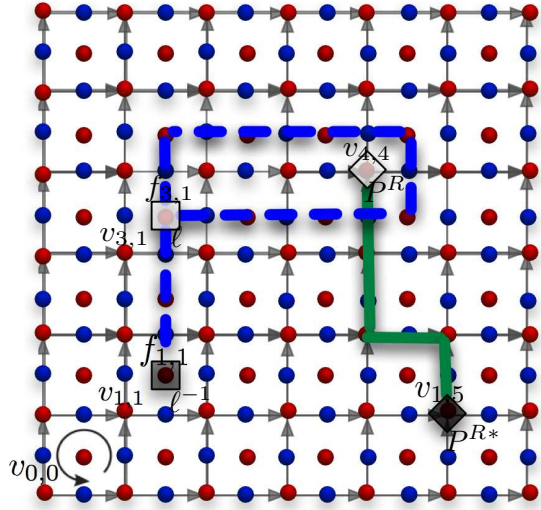}
\caption{A spin lattice model for topological order.   (a)  The system particles (blue) reside on the edges of a two complex $\Gamma$.  In a Hamiltonian formulation, all particles on edges that meet at a vertex $v$ interact as do all edges that surround a face $f$.  The orientation of edges and faces is indicated.  An ancillary set of particles (red) is placed on the vertices of the of the complex $\Gamma$ and the faces of $\Gamma$ (equivalently, the vertices of the dual complex $\tilde{\Gamma}$).  The ancillary particles on the vertices(faces) afford a handle on operations with system particles to create and guide electric(magnetic) charges which are indicated by diamonds(squares).  The braiding of one member of a vacuum magnetic charge $[\ell]$ pair around one member of a vacuum electric charge $R$ pair is shown:  $\mathcal{R}^2 \ket{0_{[\ell]};(v_{3,1},f_{3,1}),(v_{1,1},f_{1,1})}\ket{{\bf 1}^R;(v_{4,4},v_{1,5})} \rightarrow \frac{1}{\sqrt{|[\ell]|}}\sum_{\ell\in [\ell]} \ket{(P^{R_1^+},\ell);(v_{3,1},f_{3,1})}\ket{(P^{R_1^+},\ell^{-1});(v_{1,1},f_{1,1})} \ket{R(\ell);(v_{4,4},v_{1,5})} $
.}
\label{fig:1}
\end{center}
\end{figure}

We focus on a specific two-complex $\Gamma$ which is a square
lattice with boundary.  For any two complex with boundary and without
holes, there exists a ground state $\ket{GS}$ such that $H_{\rm
 TO}\ket{GS}=-(|\mathcal{V}|+|\mathcal{F}|)|\ket{GS}$ and it is
unique (for the argument see e.g. \cite{Bullock:07}).  The convention
for edge and face orientations is shown in Fig. \ref{fig:1}.  We
slightly abuse notation by labeling the particles according to
location relative to a face index $f_j$ and vertex index $v_j$ (see
Fig. \ref{fig:1}a).  For instance, a vertex ancillary particle at
vertex $v_{i,j}$ will be labeled $v_{i,j}$ and a face ancillary
particle at face $f_{i,j}$ will be labeled $f_{i,j}$.  The system
particle on edge $e=[v_{i,j},v_{k,l}]$ will be labeled $e_{i,j;k,l}$.
When we are referring to the actual spatial locations $f$ and $v$ it
will be made clear.

In Table \ref{tab:synth} we give an algorithmic procedure to prepare
the ground state of the Hamiltonian $H_{\rm TO}$ for an arbitrary
finite group $G$.  We begin with all system particles and face
ancillae in state $\ket{e}=\ket{0}$.  This guarantees that the initial
system state satisfies the zero flux condition,
i.e. $B(f)\ket{\psi}_S=\ket{\psi}_S \forall f$.  All vertex ancillae
are prepared in the state $\ket{\tilde{0}}$ where
$\ket{\tilde{j}}=\frac{1}{\sqrt{|G}}\sum_{k=0}^{|G|-1} e^{2\pi ij
 k/|G|}\ket{k}$.  In the case $G=\mathbb{Z}_2$, this algorithm
produces the ground state of the planar version of Kitaev's toric code
\cite{kitaev_toric}.  For that model all operations can be done with
qubits, and the permutation rep of the group is $L_e=R_e={\bf 1}_2$
and $L_{g_1}=R_{g_1}=\sigma^x$.  Controlled operations involve only
CNOT gates and the correction gates $Z^j=(\sigma^z)^j$.  Another
scheme for constructing this ground state using single qubit
measurements and feedforward on a prepared cluster state is given in
\cite{Han:07}.

Our algorithm has a computational depth of $O(|G|(m+n))$ but one might
wonder if a faster ground state preparation procedure is possible.
The answer is no if the initial state is uncorrelated and the
available set of operations is quasi-local.  The reason is that the
final state has global correlations that are created by quasi-local
operations.  In our algorithm these operations are measurements but
they could also be adiabatic turn on of the summands of $H_{TO}$. The
time scale to perform the quasi-local operations (here the measurement
time of $A(v)$ establishes a light cone for the flow of correlations.
In \cite{Bravyi:06} it was shown by an application of the
Lieb-Robinson bound that the minimal time to prepare a topologically
ordered state beginning in a completely uncorrelated state is of the
order of the length of the correlations.  Since the correlation length
scales as the linear dimension of our lattice, our algorithm is
essentially optimal.

\begin{table}
{\small
\begin{tabular}{lll}
\noindent
By single particle measurement, prepare the initial state $\ket{\Psi_{\rm in}}=\otimes_{e\in\mathcal{E}} \ket{0}_{e}\otimes_{v\in \mathcal{V}}\ket{\tilde{0}}_v \otimes_{f\in \mathcal{F}}\ket{0}_f$.\\
for $k=0:m-1$\\
\ \ for $j=0:n$\\
\ \ \ \  Apply the unitary $W(v_{j,k})$ where $W(v)=\sum_{h\in G}\ket{h}_{v}\bra{h}\otimes T_h(v)$. \\
\ \ \ \  Measure ancilla $v_{j,k}$ in the basis $\{\ket{\tilde{j}}\}$ (probabilty for outcome $j$ is $1/|G|$).\\
\ \ \ \   For the outcome $\ket{\tilde{j}}$ apply the single qudit operation $Z^j(e_{j,k;j,k+1})$ 
where $Z^j(e)=\sum_{k=0}^{|G|-1} e^{i 2 \pi j k/|G|}\ket{k}_{e}\bra{k}$.\\
\ \ \ \  j++ \\
\ \ end\\
\ \ k++ \\
end\\
j=0\\
for $j=0:n-1$\\
\ \ Apply the unitary $W(v_{j,k})$.\\
\ \  Measure ancilla $v_{j,k}$ in the basis $\{\ket{\tilde{j}}\}$ (probabilty for outcome $j$ is $1/|G|$).\\
\ \   For the outcome $\ket{\tilde{j}}$ apply $Z^j(e_{j,m;j+1,m})$.\\
\ \ j++ \\
end\\
\end{tabular}
\caption{
\label{tab:synth}
Algorithm: \emph{Ground state-synthesis}.  An algorithm for preparation of the ground state $\ket{GS}$ of $H_{\rm TO}$ over a finite group $G$ on an $(n+1)\times (m+1)$ square lattice with boundary satisfying
$B(f)\ket{GS}=A(v)\ket{GS}=\ket{GS} \forall v\in\mathcal{V},f\in\mathcal{F}$.  The algorithm works by beginning in a $+1$ co-eigenstate of all face operators $B(f)$ and applying vertex ancilla assisted projections $A(v)$.  Using single particle operations conditioned on measurement outcomes of vertex ancilla, this algorithm outputs the state $\ket{\Psi_{\rm out}}=\prod_{v\in \mathcal{V}\setminus v_{n,m}}[\frac{1}{\sqrt{|G|}}\sum_{h\in G}T_h(v)]\ket{\Psi_{\rm in}}$.  No projection $A_{v_{n,m}}$ is needed because for a two complex $\Gamma$ with boundary, $\prod_{v\in \mathcal{V}}A(v)\ket{e}^{\otimes |\mathcal{E}|}=\prod_{v\in \mathcal{V}\setminus v_{n,m}}A(v)\ket{e}^{\otimes |\mathcal{E}|}$.   Each controlled gate operation requires $O(|G|)$ elementary single and two qudit operations and the algorithm has complexity $O(|G|nm)$.  Since $[L_h,R_{h'}]=0$, the operators performing local gauge transformations commute: $[W(v),W(v')]=0$. Up to the last column then, all columnwise unitary operations $W(v)$ and subsequent correction gates $Z^j(e)$ can be done in parallel.  For the last column the correction gates do not commute with the operators $W(v)$ hence the operations are done serially so that the computational depth of the algorithm is $O(|G|(m+n))$.}
}
\end{table}

\section{Simulation of $\mathrm{D} ( S_3 )$}

\subsection{Ground state preparation}

The algorithm \emph{Ground state-synthesis} can be built using the
tools for single- and two-qudit gates described above.  The controlled
gauge transformations $W(v)$ are built up by applying $L_h$
or $R_{h^{-1}}$ to left,right, top, or bottom neighbors of the vertex
$v$ controlled on $v$ begin in state $\ket{h}$.  Notice that unitaries
$L_h$ and $R_h$ are maximally sparse in the logical basis, hence using
the QR decomposition for simulating unitaries \cite{Brennen:05}, each
such operation can be built using a small number two qudit diagonal
phase phase gates $U=e^{i\phi \ket{g_j}_A\bra{g_j}\otimes
 \ket{g_j}_B\bra{g_j}}$ conjugated by single qudit Givens rotations.
\footnote{Using the software package in \cite{Bullockpost} we find a
 construction for the operator $\sum_{h\in S_3}\ket{h}\bra{h}\otimes
 L_h$ or $\sum_{h\in S_3}\ket{h}\bra{h}\otimes R_h$ in $37$
 controlled phase gates.  This count may be reduced to $8$ controlled
 phase gates using a qutrit-qubit encoding for the group elements
 (see \ref{S3rep})}.  Measurement of vertex ancilla are performed in
the Fourier basis $\ket{\tilde{j}}_v$, hence a one qudit Fourier
transform $F_v$ is needed before measuring in the logical basis.  That
operator is not sparse but can be constructed with fewer than $13$
steps using parallel Givens rotations \cite{Brennen:05}.

\subsection{The particle spectrum} 

 The $8$ irreps for $\mathrm{D} ( S_3 )$ are
listed below with their corresponding quantum dimensions: \be
\begin{array}{lll}
&&\Pi^{[e]}_{R_1^+}\quad d=1 \quad {\rm (vacuum)}\\
&&\Pi^{[c]}_{\beta_0}, \Pi^{[t]}_{\gamma_0}\quad d=2,3 \quad {\rm (pure\ magnetic\ charges)} \\
&&\Pi^{[e]}_{R_1^-}, \Pi^{[e]}_{R_2}\quad d=1,2 \quad {\rm (pure\ electric\ charges)}\\
&&\Pi^{[c]}_{\beta_1}, \Pi^{[c]}_{\beta_2},\Pi^{[t]}_{\gamma_1}\quad d=2,2,3 \quad {\rm (dyonic\ combinations)}\\
\end{array}
\label{spectrum}
\ee
A complete derivation of the fusion rules for this model is given in \cite{Propitius:95}.

Magnetic charges are labeled by conjugacy classes of the the group
$G$.  Recall the conjugacy class $C_h$ is defined $C_h=\{ghg^{-1}
|g\in G\}$.  For $S_3$ we label $[e]=C_e=S_3, [c]=C_{c_+},
[t]=C_{t_0}$.  The magnetic flux across a face is given by the ordered
product of group elements represented by the basis states of the edges
surrounding the face.  The order is taken along a closed
counterclockwise cycle, beginning at an origin $v$.  Except for the
trivial flux $e$ case, the origin from which the product is taken is
also important as not labeling the origin is equivalent to only
specifying the conjugacy class or magnetic charge.  The projector onto
states with magnetic flux $\ell$ at face $f$ as computed taking a
connected counterclockwise cycle around $f$ with the origin at vertex
$v$ is:
\begin{equation}
B_{\ell}(v,f)=\sum_{\{ \prod_{e_k\in\partial f}h_k=\ell|e_0=[v,*]\}}\otimes_{e_k\in \partial f}\ket{h_k^{-o_f(e_k)}}_{e_k}\bra{h_k^{-o_f(e_k)}},
\end{equation}
The projector onto states with magnetic charge $[\ell]$ at face $f$ is
then the sum over all fluxes in the same conjugacy class: $\sum_{\ell
 \in [\ell]}B_{\ell}(v,f)$, in which case the specification of origin
is unnecessary.  The identity element $e$ is its own conjugacy class,
hence we use the convention $B_{e}(v,f)\equiv B(f)$.

Electric charges are labeled by irreps of the centralizer of a conjugacy class of $G$.
The centralizer is defined $N_h=\{g\in G | gh=hg\}$.  For $S_3$, 
we have $N_{[e]}=S_3, N_{[t]}\cong\{e,t_0\}\cong \mathbb{Z}_2, N_{[c]}=\{e,c_+,c_-\}\cong \mathbb{Z}_3$.

For a finite group $G$ every representation is equivalent to a unitary
representation.  Consider projection operators onto subspaces
belonging to the unitary irreducible representation, or unirep, $R$ with matrix
representation $R$ and dimension $|R|$:
\be
P_{\mu,
 \nu}^R=\frac{|R|}{|G|} \sum_{g\in G} [R(g)^*]_{\mu,\nu} g
\ee
This projection operator satisfies:
\[
P_{\mu,\nu}^R P_{\kappa,\lambda}^{R'} = \delta_{R,R'}\delta_{\nu,\kappa} P_{\mu,\lambda}^R.
\]
as is verified by taking the inner product over $G$,
\[
\begin{array}{lll}
P_{\mu,\nu}^R P_{\kappa,\lambda}^{R'}&=&\frac{|R||R'|}{|G|^2}\sum_{h_1,h_2\in G}[R(h_1)^*]_{\mu,\nu}[R'(h_2)^*]_{\kappa,\lambda}h_1h_2\\
&=&\frac{|R||R'|}{|G|^2}\sum_{h_1,h_2\in G}[R(h_1)^*]_{\mu,\nu}\sum_{\gamma}[R'(h_1^{-1})^*]_{\kappa,\gamma}[R'(h_1h_2)^*]_{\gamma,\lambda}h_1h_2\\
&=&\frac{|R||R'|}{|G|^2}\sum_{h'\in G}\sum_{\gamma}\Big[\sum_{h_1\in G}[R(h_1)^*]_{\mu,\nu}[R'(h_1)]_{\gamma,\kappa}\Big][R'(h')^*]_{\gamma,\lambda}h'\\
&=&\frac{|R||R'|}{|G|^2}\sum_{h'\in G}\sum_{\gamma}\frac{|G|}{|R|}\delta_{R,R'}\delta_{\kappa,\nu}\delta_{\gamma,\mu}[R(h')^*]_{\gamma,\lambda}h'\\
&=&\frac{|R|}{|G|}\delta_{R,R'}\delta_{\kappa,\nu}\sum_{h'\in G}[R(h')^*]_{\mu,\lambda}h'\\
&=&\delta_{R,R'}\delta_{\kappa,\nu}P_{\mu,\lambda}^R\\
\end{array}
\]
We can obtain a set of  $|R|$ orthonormal basis states ${\ket{R_{\lambda}}}$
which satisfy
\[
P_{\mu,\nu}^{R'} \ket{R_{\lambda}}=\delta_{R,R'}\delta_{\nu,\lambda}\ket{R_{\mu}}
\]
by applying the set ${P_{\mu,\nu}^R}$ over all $\mu$ for $\nu$ fixed onto some vector $\ket{\chi}$  with $P^R_{\nu,\nu}\ket{\chi} \neq 0$,
and normalizing.

For pure electric charges in $\mathrm{D} ( S_3 )$, we are interested in unireps of $N_{[e]}=S_3$.  There are three unireps:  the one dimensional identity rep $R_1^+$, the one dimensional signed rep $R_1^-$, and the two dimensional rep $R_2$.  The projection operators are  
\be 
P^{R_1^+}=\frac{1}{6}(e+t_0+t_1+t_2+c_++c_-)
\ee
for the identity irrep, and 
\be 
P^{R_1^-}=\frac{1}{6}(e-t_0-t_1-t_2+c_++c_-)
\ee
for the signed irrep, and:
\be
\begin{array}{lll}
P^{R_2}&=&\frac{1}{3} \Bigg[
\left(
\begin{array}{cc}
1    &  0\\
 0  & 1  \\   
\end{array}
\right) e + 
\left(
\begin{array}{cc}
0    & 1 \\
1   & 0  \\   
\end{array}
\right)  t_0 + 
\left(
\begin{array}{cc}
0    &  \xi \\
 \xi^{\ast}   & 0  \\   
\end{array}
\right)
t_1 +
\left(
\begin{array}{cc}
0    &  \xi^{\ast} \\
 \xi   & 0  \\   
\end{array}
\right)
t_2\\
&& 
+
\left(
\begin{array}{cc}
\xi^{\ast}    &  0 \\
 0   & \xi  \\   
\end{array}
\right)
c_+ +
\left(
\begin{array}{cc}
\xi    &  0 \\
 0   & \xi^{\ast}  \\   
\end{array}
\right)
 c_-\Bigg]
 \end{array}
\ee
for the $2$ dimensional irrep $R_2$ with $\xi=e^{i2\pi/3}$.
In the context of the lattice spin model, charge at a vertex $v$ will correspond to applying the components of these projection operators, with group action being local gauge transformations at vertex $v$, to some system state $\ket{\Psi}$
\be
\begin{array}{lll}
P^{R_1^+}(v)\ket{\Psi}/\sqrt{||\cdot||}&=&\frac{1}{\sqrt{6}}[T_e(v)+T_{t_0}(v)+T_{t_1}(v)+T_{t_2}(v)+
T_{c_-}(v)+T_{c_+}(v)]\ket{\Psi}\\
P^{R_1^-}(v)\ket{\Psi}/\sqrt{||\cdot||}&=&\frac{1}{\sqrt{6}}[T_e(v)-T_{t_0}(v)-T_{t_1}(v)-T_{t_2}(v)+
T_{c_-}(v)+T_{c_+}(v)]\ket{\Psi}\\
P^{R_2}_{00}(v)\ket{\Psi}/\sqrt{||\cdot||}&=&\frac{1}{\sqrt{3}} [T_e(v)+\xi^*T_{c_+}(v)+\xi T_{c_-}(v)]\ket{\Psi}\\
P^{R_2}_{01}(v)\ket{\Psi}/\sqrt{||\cdot||}&=&\frac{1}{\sqrt{3}} [T_{t_0}(v)+\xi T_{t_1}(v)+\xi^* T_{t_2}(v)]\ket{\Psi}\\
P^{R_2}_{10}(v)\ket{\Psi}/\sqrt{||\cdot||}&=&\frac{1}{\sqrt{3}} [T_{t_0}(v)+\xi^* T_{t_1}(v)+\xi T_{t_2}(v)]\ket{\Psi}\\
P^{R_2}_{11}(v)\ket{\Psi}/\sqrt{||\cdot||}&=&\frac{1}{\sqrt{3}} [T_e(v)+\xi T_{c_+}(v)+\xi^* T_{c_-}(v)]\ket{\Psi}\\

\end{array}
\ee We obtain two copies of a basis for the $2$ dimensional irrep
$R_2$, namely $\{P^{R_2}_{00}, P^{R_2}_{10}\}$ and the conjugate basis
$\{P^{R_2^*}_{00}=P^{R_2}_{11},P^{R_2^*}_{10}=P^{R_2}_{01}\}$.  The
system state $\ket{\Psi}$ here can be interpreted as the state at some
intermediate stage of the state synthesis algorithm before a gauge
symmetrization at vertex $v$ has been performed.

For the dyonic combination with flux $[c]$ the charges are labeled by unireps of the $N_{[c]}\cong\mathbb{Z}_3$.  There are three unireps of $\mathbb{Z}_3$: $R_1^0,R_1^1$, and $R_1^2$, and since it is an Abelian group, they are all one dimensional.  The projection operators for $N_{c_{\rho}}$  are  
\be 
\begin{array}{lll}
P^{R_1^0}&=&\frac{1}{3}(e+c_{\rho}+c_{-\rho}),\\
P^{R_1^1}&=&\frac{1}{3}(e+\xi c_{\rho}+\xi^{\ast} c_{-\rho}),\\
P^{R_1^2}&=&\frac{1}{3}(e+\xi^{\ast} c_{\rho}+\xi c_{-\rho})
\end{array}
\ee

For the dyonic combination with flux $[t]$ the charges are labeled by unireps of $N_{[t]}\cong \mathbb{Z}_2$.  There are two one dimensional unireps of $\mathbb{Z}_2$: $R_1^3$ and $R_1^4$.
The projection operators for $N_{t_i}$ are  
\be 
\begin{array}{lll}
P^{R_1^3}&=&\frac{1}{2}(e+t_i),\\
P^{R_1^4}&=&\frac{1}{2}(e- t_i),\\
\end{array}
\label{projNt}
\ee

Notice that  there is a relation between projection operators for the centralizer of the identity and the products $P^{R(N_{[\alpha]})}\alpha$ for $\alpha\in [\alpha]$:
\[ 
\begin{array}{lll}
P^{R_1^1}\frac{1}{2}(c_++c_-)&=&\cos(\xi) P^{R_2}_{01},\\
P^{R_1^2}\frac{1}{2}(c_++c_-)&=&\cos(\xi)P^{R_2}_{10}\\
P^{R_1^3}\frac{1}{3}(t_0+t_1+t_2)&=&P^{R_1^+},\\
P^{R_1^4}\frac{1}{3}(t_0+t_1+t_2)&=&-P^{R_1^-},\\
\end{array}
\]

In the spin lattice model electric(magnetic) charges correspond to violations of the local vertex(face) constraints and the state of a dyonic particle at vertex and face location $(v,f)$ will be denoted:  
\[
\ket{(P^R_{\mu,\nu},g);(v,f)}
\]
where $P^R_{\mu,\nu}$ labels an irrep of G as above and $g$ is the
flux at face $f$, evaluated taking a counterclockwise cycle with base
point at $v$.  For example, the state of magnetic flux $\ell$ particle
located at face $f$ and its anti-particle located at
face $f'$ (with magnetic flux evaluated with respect to the origin
$v$) is
\[
\ket{(P^{R^1_+},\ell);(v,f)}\ket{(P^{R^1_+},\ell^{-1});(v,f')}.
\]

For our model most excitations created in the bulk of the lattice
appear as particle anti-particle pairs (such that the total charge of
the pair is zero).  Although single particle excitations can be made
by creating them at the boundary.  We will describe digital
simulations of braiding of pure charges and pure fluxes.

\subsection{Anyonic dynamics}
\label{sec:braid}
Before deriving a sequence of operations to create and move anyons in
the spin lattice let's review the rules for braiding charges and
fluxes in anyonic models.  We will write $\ket{a}$ to represent a
magnetic flux of value $a$ and $\ket{a,a^{-1}}$ for a flux anti-flux
pair.  For electric charge pairs we have one charge that transforms
under the irrep $R$ and the anti-charge which transforms under the
complex conjugate representation $R^*$.  We introduce the bases
$\{\ket{\mu}_R\}_{i=0}^{|R|-1},\{\ket{\nu}_{R^*}\}_{j=0}^{|R|-1}$ on
which the representations act and write a generic state of an electric
charge anti-charge pair as a $|R|\times |R|$ matrix:
\[
\ket{M^R}=\frac{1}{\sqrt{|R|}}\sum_{\mu,\nu} M^R_{\mu,\nu}\ket{\mu}_R\otimes \ket{\nu}_{R^*}
\]
with the normalization chosen such that $\sum_{\mu,\nu}|M^R_{\mu,\nu}|^2=|R|$.

Interchanging two fluxes, a left flux $a$ and a right flux $b$ in a
counterclockwise sense is described by the action of the monodromy
operator $\mathcal{R}$:
\[
\mathcal{R}\ket{a}\ket{b}=\sigma \ket{a}\ket{aba^{-1}}=\ket{aba^{-1}}\ket{a}
\]
where $\sigma$ is the particle interchange operator.  Squaring the
monodromy operator gives the action of braiding two fluxes
\be
\mathcal{R}^2\ket{a}\ket{b}=\ket{(ab)a(ab)^{-1}}\ket{aba^{-1}}
\ee
Braiding a flux $b$ around a flux anti-flux pair $(a,a^{-1})$ is
equivalent to braiding first around one then around the other (we can
order the particles left to right $(1,2,3)$)
\begin{equation}
\mathcal{R}^2_{1,2}\otimes \mathcal{R}^2_{1,3}\ket{b}\ket{a,a^{-1}}=\ket{b}\ket{bab^{-1},ba^{-1}b^{-1}}
\end{equation}
\label{fluxpairbraid}
If $\ket{bab^{-1},ba^{-1}b^{-1}}\neq \ket{a,a^{-1}}$ then we say that
$\ket{a,a^{-1}}$ has magnetic charge.  For each conjugacy class
$[\ell]$, there is one unique \emph{chargeless} state defined:
\be
\ket{0_{[\ell]}}=\frac{1}{\sqrt{|[\ell]|}}\sum_{\ell\in
 [\ell]}\ket{\ell,\ell^{-1}}
\label{zmcharge}
\ee

Electric charges moving past each other have no effect, only
the braiding of fluxes around charges has an effect.
Specifically, if we braid a flux $\ket{h}$ around one electric charge
in the pair $\ket{M^R}$ we obtain:
\begin{equation}
\mathcal{R}^2_{1,2}\ket{h}\ket{M^R}=\ket{h}\ket{R(h)M^R}
\end{equation}
and if we braid around the anti-charge we obtain:
\begin{equation}
\mathcal{R}^2_{1,3}\ket{h}\ket{M^R}=\ket{h}\ket{M^RR(h^{-1})}
\end{equation}
where $R(h)$ is the matrix representation $R$ of the group element $h$.  Braiding around both charges is a conjugation
\[
\mathcal{R}^2_{1,2}\otimes\mathcal{R}^2_{1,3}\ket{h}\ket{M^R}=\ket{h}\ket{R(h)M^RR(h^{-1})}
\]
For each irrep $R$ there is one unique \emph{fluxless} state that is invariant under conjugation:
\begin{equation}
\ket{{\bf 1}_{|R|}}=\frac{1}{\sqrt{|R|}}\sum_{\mu}\ket{\mu}_R\otimes \ket{\mu}_{R^*}.
\label{elecvac}
\end{equation}

\subsubsection{Magnetic charges in $\mathrm{D} ( S_3 )$}
\label{mcharge}
Consider the creation of the chargeless magnetic flux pair for the conjugacy class
$[\ell]$
\be
\ket{0_{[\ell]};(v_{i+1,j+1},f_{i,j}),(v_{i+1,j+1},f_{i,j+1})}=\frac{1}{\sqrt{|[\ell]|}}\sum_{\ell\in[\ell]}\ket{(P^{R_1^+},\ell^{-1});(v_{i+1,j+1},f_{i,j})}\ket{(P^{R_1^+},\ell);(v_{i+1,j+1},f_{i,j+1})}
\ee
This state can be created by starting in the ground state $\ket{GS}$
and acting on one edge which is a shared boundary of the two faces:
\[
\ket{0_{[\ell]};(v_{i+1,j+1},f_{i,j}),(v_{i+1,j+1},f_{i,j+1})}=\frac{1}{\sqrt{|[\ell]|}}\sum_{\ell\in[\ell]}R_{\ell}(e_{i,j+1;i+1,j+1})\ket{GS}
\]
Note that the right multiplication operator
$R_{\ell}(e_{i,j+1;i+1,j+1})$ commutes with all vertex operators
except $A(v_{i+1,j+1})$.  However, the sum of $R_{\ell}$ over all
elements of the conjugacy class does commute with it.  To reduce
clutter here, we write $v=v_{i+1,j+1}$ and $e=e_{i,j+1;i+1,j+1}$,
then:
\[\fl
\begin{array}{lll}
\sum_{\ell\in[\ell]}  A(v)R_{\ell}(e) A(v)&=&\frac{1}{|G|^2}\sum_{g,g'}\tilde{T}_g(v)\tilde{T}_{g'}(v)\otimes[\sum_{\ell\in[\ell]} R_{g^{-1}}(e)R_{\ell}(e)R_{g'^{-1}}(e)]\\
&=&\frac{1}{|G|^2}\sum_{g,g'}\tilde{T}_g(v)\tilde{T}_{g'}(v)\otimes[\sum_{\ell\in[\ell]} R_{g^{-1}}(e)R_{g^{-1}\ell g}(e)R_{g'^{-1}}(e)]\\
&=&\frac{1}{|G|^2}\sum_{g,g'}\tilde{T}_g(v)\tilde{T}_{g'}(v)\otimes[\sum_{\ell\in[\ell]} R_{\ell}(e)R_{g^{-1}}(e)R_{g'^{-1}}(e)]\\
&=&\frac{1}{|G|^2}\sum_{\ell\in[\ell]} R_{\ell}(e)\sum_{g,g'}T_g(v)T_{g'}(v)\\
&=&\sum_{\ell\in[\ell]} R_{\ell}(e) A(v)^2\\
&=&\sum_{\ell\in[\ell]} R_{\ell}(e) A(v)
\end{array}
\]
where $\tilde{T}_{g}(v)R_{g^{-1}}(e)\equiv T_{g}(v)$, since $A(v)\ket{GS}=\ket{GS}$.  Therefore, only the face constraints $B(f_{i,j}),B(f_{i,j+1})$ are violated.
Starting from the vacuum (ground) state $\ket{GS}$, this state is created by preparing the ancilla $f_{i,j}$ in the state 
$\ket{0_{[\ell ]}}_{f_{i,j}}$ where
$\ket{0_{[\ell]}}=\frac{1}{\sqrt{[\ell]}}\sum_{\ell\in [\ell]}\ket{\ell}$, applying the two qudit unitary
\[
F_{[\ell]}(f_{i,j})={\bf 1}_{|G|-|[\ell]|}\otimes {\bf 1}_{|G|}+\sum_{\ell\in [\ell]}\ket{\ell}_{f_{i,j}}\bra{\ell}\otimes R_{\ell}(e_{i,j+1;i+1,j+1})
\]
and measuring the face ancilla in the basis $\{\ket{k_{[\ell]}}=Z_{|[\ell]|}^k\ket{0_{[\ell]}}\}_{k=0}^{|[\ell]|-1}$, where $Z_{[\ell]}^k=\sum_{\ell_m\in[\ell]}e^{i2\pi k m/|[\ell]|}\ket{\ell_m}\bra{\ell_m}$ (where
we have labeled the group elements in $[\ell]=\{l_0,\ldots, \ell_{|[\ell]|-1}\}$).  For the outcome $0_{[\ell]}$ the target magnetic charge state is created. Otherwise for outcome $k_{[\ell]}$, we need a correction step.  To do this prepare the ancilla $f_{i,j}$ in the state $\ket{e}_{f_{i,j}}$.  Apply the controlled operation
$\Lambda(v_{i+1,j+1},f_{i,j})$ where
\[
\Lambda(v,f)= \sum_{g\in G}B_{g}(v,f)\otimes L_{g}(f).
\]
which maps the ancilla $f$ to state $\ket{g}_f$ when the flux at $f$ (with respect to the base point $v$) is $g$.  Such a controlled operation can be decomposed into elementary two qudit controlled rotation operators with each edge $e_k$ surrounding $f$ as a control and the ancilla as the target, viz.
\[
\begin{array}{lll}
\Lambda(v_{i+1,j+1},f_{i,j})&=&    \Lambda_{\rm right}(f_{i,j})\Lambda_{\rm bottom}(f_{i,j})\Lambda_{\rm left}(f_{i,j})\Lambda_{\rm top}(f_{i,j}) \\
\Lambda(v_{i+1,j},f_{i,j})&=&\Lambda_{\rm top}(f_{i,j})\Lambda_{\rm right}(f_{i,j})\Lambda_{\rm bottom}(f_{i,j})\Lambda_{\rm left}(f_{i,j})  \\
\Lambda(v_{i,j},f_{i,j})&=&\Lambda_{\rm left}(f_{i,j})\Lambda_{\rm top}(f_{i,j})\Lambda_{\rm right}(f_{i,j})\Lambda_{\rm bottom}(f_{i,j})  \\
\Lambda(v_{i,j+1},f_{i,j})&=&\Lambda_{\rm bottom}(f_{i,j})\Lambda_{\rm left}(f_{i,j})\Lambda_{\rm top}(f_{i,j})\Lambda_{\rm right}(f_{i,j})  \\
\end{array}
\]
where
\[
\begin{array}{lll}
\Lambda_{\rm right}(f_{i,j})&=&[\sum_{g\in G}\ket{g}_{e_{i,j+1;i+1,j+1}}\bra{g}\otimes L_{g^{-1}}(f_{i,j})],\\
\Lambda_{\rm bottom}(f_{i,j})&=&[\sum_{g\in G}\ket{g}_{e_{i,j;i,j+1}}\bra{g}\otimes L_{g^{-1}}(f_{i,j})]\\
\Lambda_{\rm left}(f_{i,j})&=&[\sum_{g\in G}\ket{g}_{e_{i,j;i+1,j}}\bra{g}\otimes L_{g}(f_{i,j})],\\
\Lambda_{\rm top}(f_{i,j})&=&[\sum_{g\in G}\ket{g}_{e_{i+1,j;i+1,j+1}}\bra{g}\otimes L_{g^{-1}}(f_{i,j})].\\
\end{array}
\]
The convention here is that one takes the product of controlled left multiplication operations on the face ancilla along a counterclockwise path around $f$ beginning at vertex $v$ where the operator applied to $f$ is 
$L_{g^{\mp 1}}$ depending on the orientation of the edge relative to the face.
Next apply the single qudit phase gate $[{\bf 1}_{|G|-|[\ell]|}\oplus Z_{|[\ell]|}^{k}](f_{i,j})$
and finally apply $\Lambda(f_{i,j})^{-1}$ to disentangle the ancilla from the system.

We now describe in detail how to move magnetic charges from one face $f$ to an adjacent face $f'$.  Essentially it involves coherently mapping the value of flux at $f$ to the face ancilla $f$ and applying a controlled operation on an edge $e\in\partial f\sqcup\partial f'$ (the shared boundary of the faces $f,f'$).  After this controlled operation the face ancilla $f$ is disentangled from the system by mapping the flux at face $f'$ to ancilla $f'$ and performing a controlled operation between ancillae $f,f'$ and finally reversing the mapping on $f'$.  In this protocol we are careful to demand only single qudit and nearest neighbor two qudit interactions.

Here is the protocol to move a magnetic flux one face unit to the right:
\[
\ket{(P^{R_1^+},x);(v_{i+1,j+1},f_{i,j})}\rightarrow\ket{(P^{R_1^+},x);(v_{i+1,j+1},f_{i,j+1})}.
\]
\begin{itemize}
\item
Prepare the face ancillae $f_{i,j},f_{i,j+1}$ in state $\ket{e}$
\item
Coherently map the state of the flux at face $f_{i,j}$ to the ancilla $f_{i,j}$ via $\Lambda(v_{i+1,j+1},f_{i,j})$
\item Apply the controlled unitary $Y(f_{i,j}, e_{i,j+1;i+1,j+1})$ where we define
\[
Y(f,e)=\left\{\begin{array}{c}\sum_{h\in G}\ket{h}_{f}\bra{h}\otimes R_{h}(e)\quad o_f(e)=+1  \\ \sum_{h\in G}\ket{h}_{f}\bra{h}\otimes L_{h^{-1}}(e)\quad o_f(e)=-1 \end{array}\right.
\]
The remaining steps disentangle the ancilla from the system.
\item  Map the flux value at the face $f_{i,j+1}$ to the face ancilla $f_{i,j+1}$ by applying $\Lambda(v_{i+1,j+1},f_{i,j+1})$
\item Swap qudits $e_{i,j+1;i+1,j+1}$ and $f_{i,j+1}$
\item Apply the unitary $u=\sum_{g\in G}\ket{g}_{f_{i,j+1}}\bra{g}\otimes L_{g^{-1}}(f_{i,j})$
\item Swap qudits $e_{i,j+1;i+1,j+1}$ and $f_{i,j+1}$
\item Apply $\Lambda(v_{i+1,j+1},f_{i,j+1})^{-1}$. 
\end{itemize}
This entire process
respects superpositions over flux states and can therefore be used to propagate magnetic charges around the lattice.  A simple adaptation allows a magnetic flux move one face unit to the left:
\[
\ket{(P^{R_1^+},x);(v_{i,j},f_{i,j})}\rightarrow\ket{(P^{R_1^+},x);(v_{i,j},f_{i,j-1})}.
\]
\begin{itemize}
\item
Prepare the face ancillae $f_{i,j},f_{i-1,j-1}$ in state $\ket{e}$
\item
Coherently map the state of the flux at face $f_{i,j}$ to the ancilla $f_{i,j}$ via $\Lambda(v_{i,j},f_{i,j})$
\item Apply the controlled unitary $Y(f_{i,j}, e_{i,j;i+1,j})$.
\item  Map the flux value at the face $f_{i,j-1}$ to the face ancilla $f_{i,j-1}$ by applying $\Lambda(v_{i,j},f_{i,j-1})$
\item Swap qudits $e_{i,j;i+1,j}$ and $f_{i,j-1}$
\item Apply the unitary $u=\sum_{g\in G}\ket{g}_{f_{i,j-1}}\bra{g}\otimes L_{g^{-1}}(f_{i,j})$
\item Swap qudits $e_{i,j;i+1,j}$ and $f_{i,j-1}$
\item Apply $\Lambda(v_{i,j},f_{i,j-1})^{-1}$. 
\end{itemize}
Similarly, to move a magnetic flux one face unit down:
\[
\ket{(P^{R_1^+},x);(v_{i,j+1},f_{i,j})}\rightarrow\ket{(P^{R_1^+},x);(v_{i,j+1},f_{i-1,j})}.
\]
\begin{itemize}
\item
Prepare the face ancillae $f_{i,j},f_{i-1,j}$ in state $\ket{e}$
\item
Coherently map the state of the flux at face $f_{i,j}$ to the ancilla $f_{i,j}$ via $\Lambda(v_{i,j+1},f_{i,j})$
\item Apply the controlled unitary $Y(f_{i,j}, e_{i,j;i,j+1})$.
\item  Map the flux value at the face $f_{i-1,j}$ to the face ancilla $f_{i-1,j}$ by applying $\Lambda(v_{i,j+1},f_{i-1,j})$
\item Swap qudits $e_{i,j;i,j+1}$ and $f_{i-1,j}$
\item Apply the unitary $u=\sum_{g\in G}\ket{g}_{f_{i-1,j}}\bra{g}\otimes L_{g^{-1}}(f_{i,j})$
\item Swap qudits $e_{i,j;i,j+1}$ and $f_{i-1,j}$
\item Apply $\Lambda(v_{i,j+1},f_{i-1,j})^{-1}$. 
\end{itemize}
Finally, to move a magnetic flux one face unit up:
\[
\ket{(P^{R_1^+},x);(v_{i+1,j},f_{i,j})}\rightarrow\ket{(P^{R_1^+},x);(v_{i+1,j},f_{i+1,j})}.
\]
\begin{itemize}
\item
Prepare the face ancillae $f_{i,j},f_{i+1,j}$ in state $\ket{e}$
\item
Coherently map the state of the flux at face $f_{i,j}$ to the ancilla $f_{i,j}$ via $\Lambda(v_{i+1,j},f_{i,j})$
\item Apply the controlled unitary $Y(f_{i,j}, e_{i+1,j;i+1,j+1})$.
\item  Map the flux value at the face $f_{i+1,j}$ to the face ancilla $f_{i+1,j}$ by applying $\Lambda(v_{i+1,j},f_{i+1,j})$
\item Swap qudits $e_{i+1,j;i+1,j+1}$ and $f_{i+1,j}$
\item Apply the unitary $u=\sum_{g\in G}\ket{g}_{f_{i+1,j}}\bra{g}\otimes L_{g^{-1}}(f_{i,j})$
\item Swap qudits $e_{i+1,j;i+1,j+1}$ and $f_{i+1,j}$
\item Apply $\Lambda(v_{i+1,j},f_{i+1,j})^{-1}$. 
\end{itemize}
Fusion of a magnetic charge particle anti-particle pair can be measured
by using controlled operations to bring the constituent charges in
conjugacy class $[\ell]$ adjacent to one another at faces $(f,f')$ with shared edge $e$ and
applying first the operator $\Lambda(v,f)= \sum_{g\in G}B_{g}(v,f)\otimes
L_{g}(f)$ followed by the operator $\sum_{g\in G}\ket{g}_{f}\bra{g}\otimes L_{g^{-1}}(e)$ then measurement of the ancilla $f$ in the basis
$\ket{g_{[\ell ]}}_{f}$.  The probability to obtain the outcome
$0_{[\ell]}$ equals the probability for the pair to fuse into the
vacuum.  


\subsubsection{Electric charges in $\mathrm{D} ( S_3 )$}
\label{echarge}
Consider the following electric charge $R(N([e])$ particle at
$v_{i,j}$ and its anti-particle pair at $v_{i,j+1}$
\begin{equation}
\ket{(P^R_{\mu,\beta},e);(v_{i,j},-)}\ket{(P^{R*}_{\nu,\beta},e);(v_{i,j+1},-)},
\label{chargest}
\end{equation}
which is a labeling of the state by basis states of the irrep $R$ and
it's conjugate $R^*$.  The index $\beta$ labels different copies
of higher dimensional irreps and for physical states we will sum over the copies.
A generic state of an electric charge-anti charge pair at vertices
$(v,v')$ will then be represented as a matrix
\[
\ket{M^R;(v,v')}=\frac{1}{\sqrt{|R|}}\sum_{\mu,\nu}M^R_{\mu,\nu}\Big(\frac{1}{\sqrt{|R|}}\sum_{\beta=0}^{|R|-1}\ket{(P^R_{\mu,\beta},e);(v,-)}\ket{(P^{R*}_{\nu,\beta},e);(v',-)}\Big) \]
with the normalization chosen so that $\sum_{\mu,\nu=0}^{|R|-1}|M^R_{\mu,\nu}|^2=|R|$.  Let's see how such a state arises in this spin lattice model.
Recall that charge states are obtained by acting on a state $\ket{\Psi}$ at an intermediate stage of ground state synthesis where the gauge symmetrization has been performed at all vertices except at  $v_{i,j},v_{i,j+1}$.  Applying the following joint charge projection operators onto $\ket{\Psi}$ gives
\begin{equation}
\begin{array}{lll}
&&P^R_{\mu,\beta}(v_{i,j})P^{R*}_{\nu,\beta}(v_{i,j+1})\ket{\Psi}/\sqrt{||\cdot ||}\\
&=&\frac{|R|^2}{|G|^2}\sum_{h,h'\in G} [R(h)^{\ast}]_{\mu,\beta}[R(h')]_{\nu,\beta}T_{h}(v_{i,j})T_{h'}(v_{i,j+1})\ket{\Psi}/\sqrt{||\cdot ||}\\
&=&\frac{|R|^2}{|G|^2}\sum_{h\in G}\sum_{\gamma}\Big[\sum_{h'\in G} [R(hh'^{-1})^{\ast}]_{\mu,\gamma}[R(h')^{\ast}]_{\gamma,\beta}[R(h')]_{\nu,\beta}  T_{hh'^{-1}}(v_{i,j},v_{i,j+1})\ket{\Psi}/\sqrt{||\cdot ||}\\
&=&\frac{|R|}{|G|}\sum_{hh'^{-1}\in G}\sum_{\gamma}[R(hh'^{-1})^{\ast}]_{\mu,\gamma}\delta_{\gamma,\nu}  T_{hh'^{-1}}(v_{i,j},v_{i,j+1})\ket{\Psi}/\sqrt{||\cdot ||}\\
&=&\frac{|R|}{|G|}\sum_{hh'^{-1}\in G}[R(hh'^{-1})^{\ast}]_{\mu,\nu} T_{hh'^{-1}}(v_{i,j},v_{i,j+1})\ket{\Psi}/\sqrt{||\cdot ||}\\
\end{array}
\end{equation}
The operator
$T_{hh'^{-1}}(v_{i,j},v_{i,j+1})=T_{h}(v_{i,j})T_{h'}(v_{i,j+1})$ acts
as $L_hR_{h'^{-1}}(e_{i,j;i,j+1})$ on the connecting edge.  Before the
projections, the system particle $e_{i,j;i,j+1}$ on the edge
connecting the vertices is in state $\ket{e}$, while after the local
gauge transformations on the boundaries it is in the state
$\ket{hh'^{-1}}$.  Hence the charge state Eq. \ref{chargest} can
equivalently be obtained by beginning in the fully gauge invariant
ground state $\ket{GS}$ and applying a projection operator that
depends on the local state of the connecting edge $e_{i,j;i,j+1}$:
\begin{equation}
\ket{(P^R_{\mu,\beta},e);(v_{i,j},-)}\ket{(P^{R*}_{\nu,\beta},e);(v_{i,j+1},-)}=\sqrt{|R|}\Big[\sum_{g\in G}\ket{g}_{e_{i,j;i,j+1}}\bra{g}  [R(g)^{\ast}]_{\mu,\nu}\Big]\ket{GS}.
\label{projelec}
\end{equation}
This argument in fact extends to electric charge anti-charge pairs separated by longer chains.  Consider the chain $[v_1,v_2,\ldots v_k]$ and an initial state $\ket{\Psi}$ which is gauge symmetrized over all vertices but those in the chain .  Applying the projections onto the charge pair at the boundaries and onto charge zero for the vertices $\{v_2,v_3,\ldots v_{k-1}\}$, we have
\[
\frac{P^R_{\mu,\beta}(v_1)P^{R*}_{\nu,\beta}(v_k)\ket{\Psi}}{\sqrt{||\cdot ||}}=\frac{\frac{|R|}{|G|}\sum_{h_1h_k^{-1}\in G}[R(h_1h_k^{-1})^{\ast}]_{\mu,\nu} T_{h_1h_{k}^{-1}}(v_1,v_k)\sum_{h_2,\ldots h_{k-1}}\prod_{j=2}^{k-1}T_{h_j}(v_j)\ket{\Psi}}{\sqrt{||\cdot ||}}\\
\]
The gauge transformations act on a state for the chain as
\[
\begin{array}{lll}
&&T_{h_1h_{k}^{-1}}(v_1,v_k)\sum_{h_2,\ldots h_{k-1}}\prod_{j=2}^{k-1}T_{h_j}(v_j)\ket{\ell_1}_{[v_1,v_2]}\ket{\ell_2}_{[v_2,v_3]}\ldots \ket{\ell_{k-1}}_{[v_{k-1},v_k]}\\
&=&\sum_{h_2,\ldots h_{k-1}}\ket{h_1\ell_1h_2^{-1}}_{[v_1,v_2]}\ket{h_2\ell_2h_3^{-1}}_{[v_2,v_3]}\ldots \ket{h_{k-1}\ell_{k-1}h_k^{-1}}_{[v_{k-1},v_k]}.
\end{array}
\]
It is then the product of the group elements $\ell_1\ell_2\ldots \ell_{k-1}$ along the edges of the chain that is an invariant under local gauge transformations on the vertices (excluding the boundaries).  Notice that if one of the edges had opposite orientation to the chain, say $e=[v_j,v_{j-1}]$, then the invariant would be $\ell_1\ldots \ell_{j-1} \ell_j^{-1}\ell_{j+1}\ldots \ell_{k-1}$.   Hence we find that an arbitrary state of an electric charge pair along the chain $[v_1,\ldots ,v_k]$ spanned by the edges $\{e_j\}$ is

\begin{equation}
\fl
\ket{(P^R_{\mu,\beta},e);(v_1,-)}\ket{(P^{R*}_{\nu,\beta},e);(v_k,-)}=\sqrt{|R|}\Big[\sum_{h_1,\ldots, h_{k-1}\in G}\otimes _{j=1}^{k-1}\ket{h_j}_{e_j}\bra{h_j}  [R(\prod_{j=1}^{k-1}h_j^{o(e_j)})^{\ast}]_{\mu,\nu}\Big]\ket{GS},
\end{equation}
where $o(e_j)=1$ if $e_j=[v_j,v_{j+1}]$ and $o(e_j)=-1$ if $e_j=[v_{j+1},v_j]$.

We now describe how to construct the charge state of Eq. \ref{chargest}.  This state does not violate any face constraints (since the local gauge transformations do not change magnetic flux), but it does violate the vertex constraints $A(v_{i,j}),A(v_{i,j+1})$.  To create this state, first prepare the vertex ancilla $v_{i,j}$ in state $\ket{e}_{v_{i,j}}$.  Apply the conditional unitary $K(v_{i,j},e_{i,j;i,j+1})$ defined by
\[
K(v,e)=\left\{\begin{array}{c}\sum_{g\in G}\ket{g}_{e}\bra{g}\otimes R_{g}(v)\quad e=[v,*]  \\ \sum_{g\in G}\ket{g}_{e}\bra{g}\otimes R_{g^{-1}}(v)\quad e=[*,v] \end{array}\right.
\]
and measure ancilla $v_{i,j}$ in the basis
\[
\ket{R_{\mu,\nu}}=\sqrt{\frac{|R|}{|G|}}\sum_{g\in G}[R(g)^{\ast}]_{\mu,\nu}\ket{g}
\]
The orientation of the edge dictates the left or right action taken by the operator $K$ on the vertex ancilla.

The above method only works probabilistically with probability $1/|G|$.  It is possible to prepare the vacuum electric charge states with unit probability as we know describe.  For Eq. \ref{elecvac}, the vacuum electric charge states at the boundaries of the edge $e_{i,j;i,j+1}$ are
\be
\ket{{\bf 1}_{|R|};(v_{i,j},v_{i,j+1})}=\frac{1}{|R|}\sum_{\mu,\beta}\ket{(P^R_{\mu,\beta},e);(v_{i,j},-)}\ket{(P^{R*}_{\mu,\beta},e);(v_{i,j+1},-)}.
\ee
\label{chargesym}
Using Eq. \ref{projelec} this can be rewritten as the effect of a single qudit edge operator acting on the ground state,
\be
\begin{array}{lll}
\ket{{\bf 1}_{|R|};(v_{i,j},v_{i,j+1})}&=&\sum_{\mu}\sum_{g\in G}\ket{g}_{e_{i,j;i,j+1}}\bra{g} [R(g)^{\ast}]_{\mu,\mu}\ket{GS}\\
&=&W_R(e_{i,j;i,j+1})\ket{GS}
\end{array}
\ee
where 
\[
W_R(e)=\sum_{g\in G}\ket{g}_e\bra{g} \chi^{\ast}_R(g).
\]
These operators obey the sum rule,  $\sum_R |R| W_R=|G|\ket{e}\bra{e}$.  For $S_3$ we have explicitly,
\begin{equation}
\begin{array}{lll}
W_{R_1^+}&=&{\bf 1}_6\\
W_{R_1^-}&=&\ket{e}\bra{e}+\ket{c_+}\bra{c_+}+\ket{c_-}\bra{c_-}-\ket{t_0}\bra{t_0}-\ket{t_1}\bra{t_1}-\ket{t_2}\bra{t_2}\\
W_{R_2}&=&2\ket{e}\bra{e}-\ket{c_+}\bra{c_+}-\ket{c_-}\bra{c_-}
\end{array}
\end{equation}
\label{echargeops}
Notice, that for the one dimensional irreps, the electric charge creation operator is a diagonal unitary acting on the edge.  The operator $W_{R_2}$ is not unitary but we can construct it using adaptive measurements.  First prepare the vertex ancilla $v_{i,j}$ in the state $\ket{+}_{v_{i,j}}$ where $\ket{\pm}=(\ket{0}\pm\ket{1})\sqrt{2}$ and apply the controlled unitary operation $\ket{0}_{v_{i,j}}\bra{0}\otimes U_0(e_{i,j;i,j+1})+\ket{1}_{v_{i,j}}\bra{1}\otimes U_1(e_{i,j;i,j+1})$ where 
$U_0={\rm diag}(1,1,\xi,\xi^{\ast},\xi,\xi^{\ast}), U_1={\rm diag}(1,-1,\xi^{-1/2},\xi^{1/2},\xi^{\ast},\xi)$ in the basis $\{\ket{e},\ket{t_0},\ket{t_1},\ket{t_2},\ket{c_+},\ket{c_-}\}$.  Next measure the ancilla in the basis $\ket{\pm}_{v_{i,j}}$.  The outcomes $\pm 1$ occur with probability $p_{\pm}=\frac{1}{2}\mp\frac{1}{4}$.  For the outcome $+1$, the operator $W_{R_2}$ is successfully applied and the vacuum charge state is created. Otherwise we must correct.  One strategy is to first apply the operator $U_2(e_{i,j;i,j+1})$ where $U_2={\rm diag}(1,1,\xi^{\ast},\xi,i,-i)$.  This corresponds to a combined action on the system ${\rm diag}(0,1,1,1,-\frac{\sqrt{3}}{2},-\frac{\sqrt{3}}{2})\ket{GS}$.  Now, re-prepare the ancilla in the state $\ket{e}_{v_{i,j}}$ and entangle the edge with it by the operation $X=\sum_{j=0}^2 \ket{t_j}_{e_{i,j;i,j+1}}\bra{t_j}\otimes R_{t_j}(v_{i,j})+{\bf 1}_3\otimes {\bf 1}_6$.   Next, apply a controlled gauge transformation at vertex $v_{i,j}$ that depends on the state of the edge via the operator $Y=\sum_{j=0}^2 \ket{t_j}_{e_{i,j;i,j+1}}\bra{t_j}\otimes T_{t_j^{-1}}(v_{i,j})+{\bf 1}_3\otimes {\bf 1}_6$.  This acts to map the three components of the code wavefunction with edge states in $\ket{t_j}$ to the state $\ket{e}$.  Since all these operations preserve the zero flux condition on neighboring faces, the states of the remaining lattice spins are disentangled from the $\ket{t_j}$ components of the ancilla.  Finally, we need to disentangle the ancilla from the lattice spins.  This can be realized by first applying the single qudit operation on the ancilla $U_3=(\ket{e}(\bra{t_0}+\bra{t_1}+\bra{t_2})\sqrt{3}+\ket{t_1}(\bra{t_0}+\xi\bra{t_1}+\xi^{\ast}\bra{t_2})\sqrt{3}+\ket{t_2}(\bra{t_0}+\xi^{\ast}\bra{t_1}+\xi\bra{t_2})\sqrt{3}+\ket{t_0}\bra{e})\oplus{\bf 1}_2$ and then measuring the ancilla in the Fourier basis $\ket{\tilde{k}}_{v_{i,j}}$.  For outcome $k$ apply the correction gate $Z^k(e_{i,j;i,j+1})$.  This then realizes $W_{R_2}(e_{i,j;i,j+1})\ket{GS}$.

One can prepare states with electric charge particle anti-particle pairs that are separated further than one edge, e.g.$\ket{(P^R_{\mu,0},e);(v_{i,j},-)}\ket{(P^{R*}_{\nu,0},e);(v_{i,j+2},-)}$ as follows.
Prepare vertex ancilla $v_{i,j}$  in state $\ket{e}_{v_{i,j}}$, apply the conditional unitary $K(v_{i,j},e_{i,j;i,j+1})$.  Swap the qudits at locations $v_{i,j}$ and $e_{i,j;i,j+1}$ and then swap qudits at locations $e_{i,j;i,j+1}$ and $v_{i,j+1}$.  Apply the conditional unitary $K(v_{i,j+1},e_{i,j+1;i,j+2})$ and  measure $v_{i,j}$ in the basis $\{\ket{R'_{\mu,\nu}}\}$. To deterministically prepare the vacuum state $\ket{{\bf 1}_{|R|};(v_{i,j},v_{i,j+2})}$, apply the same sequence of swaps but instead of measuring the ancilla, apply the gate $W_R$ to the ancilla and then invert the sequence of conditional unitaries and swaps.  This protocol easily extends to preparing arbitrarily long separated electric charge pairs provided care is taken during the controlled flip operations on the ancilla to keep track of the orientation of the edge for the system control.  In the case of the non-unitary gate $W_{R_2}$, another ancilla should be used to perform the operations on the ancilla which carries the information of the accumulated product of group elements along the path between the boundary vertices where the charges will be created. 

To move an electric charge from a vertex $v$ to a neighboring vertex $v'$ consider that the edges are oriented $[*,v],[v,v']$.  Then we need to coherently map the state of the edge $[*,v]$ to the product of the states of $[*,v]$ and $[v,v']$.  This is achieved by the following operator:
\[
A(v)\sum_{g,g'\in G}T_{g'}(v')T_{g}(v)\ket{g}_{[*,v]}\bra{g}\otimes \ket{g'}_{[v,v']}\bra{g'}
\]
The last projector $A(v)$ reinforces the gauge symmetry at $v$ so that total charge is conserved.  More concretely, say we wish to move a electric charge one unit to the right:
\[
\ket{(P^{R}_{\mu,\beta},e);(v_{i,j},-)}\rightarrow \ket{(P^{R}_{\mu,\beta},e);(v_{i,j+1},-)}\
\]
then the following protocol will work:
\begin{itemize}
\item
Prepare the vertex ancillae $v_{i,j},v_{i,j+1}$ in state $\ket{e}$
\item
Coherently map the state at edge $e_{i,j-1;i,j}$ to vertex ancilla $v_{i,j}$ using the operator $K^{-1}(v_{i,j},e_{i,j-1;i,j})$ and similarly for the other edge using $K^{-1}(v_{i,j+1},e_{i,j;i,j+1})$. 
\item Apply the controlled gauge transformations $W(v_{i,j})$ and $W(v_{i,j+1})$.
\item  Disentangle the ancilla $v_{i,j}$ by applying $K(v_{i,j},e_{i,j;i,j+1})$.
\item Prepare the vertex ancilla $v_{i,j}$ in state $\ket{\tilde{0}}_{v_{i,j}}$.
\item Apply the controlled gauge transformation $W(v_{i,j})$.
\item Measure vertex ancilla $v_{i,j}$ in basis $\{\ket{\tilde{j}}\}$.
\item For the outcome $\ket{\tilde{j}}$ apply the single qudit operation $Z^j(e_{i,j-1;i,j})$.
\item Disentangle the ancilla $v_{i,j+1}$ by measuring in the basis $\ket{R_{\mu,\nu}}_{v_{i,j+1}}$.
\end{itemize}
Moving charges in other directions is straightforward keeping in mind that if the edge orientations are reversed then the inverse gauge transformations should be applied.  

Finally, fusion of an electric charge pair $\ket{M^R;(v,v')}$ is realized by 
moving the charges until they overlap at one vertex (say $v'$).
The outcome of measurement on the ancilla $v'$ in the transport steps above determines the residual charge.  For outcome $\ket{R_1^+}=\ket{\tilde{0}}$ the charges are perfectly annihilated into the vacuum,
otherwise there is some residual charge.


\subsubsection{Dyons in $\mathrm{D} ( S_3 )$}
We briefly mention how to create dyonic excitations without discussing how to move them.  Recall from Eq. \ref{spectrum}, that there are three dyonic particles in $\mathrm{D} ( S_3 )$.  The explicit representations of the dyonic particle antiparticle pairs located at $(v_{i,j},f_{i,j})$ and $(v_{i,j+1},f_{i-1,j})$ are: 
\begin{equation}
\begin{array}{lll}
&&\frac{1}{\sqrt{2}}\sum_{\rho=\pm}\ket{(P^{R^1_1},c_{\rho});(v_{i,j},f_{i,j})}\ket{(P^{R^{*1}_1},c_{\rho}^{-1});(v_{i,j+1},f_{i-1,j})},\\
&&\frac{1}{\sqrt{2}}\sum_{\rho=\pm}\ket{(P^{R^2_1},c_{\rho});(v_{i,j},f_{i,j})}\ket{(P^{R^{*2}_1},c_{\rho}^{-1});(v_{i,j+1},f_{i-1,j})},\\
&&\frac{1}{\sqrt{3}}\sum_{j=0}^2 \ket{(P^{R^4_1},t_j);(v_{i,j},f_{i,j})}\ket{(P^{R^{*4}_1},t_j);(v_{i,j+1},f_{i-1,j})}.\\
\end{array}
\end{equation}
To create these states first create the vacuum magnetic charge states
$\ket{0_{[\ell]};(v_{i,j},f_{i,j}),(v_{i,j+1},f_{i-1,j})}$ with $[c]$ for the first two and $[t]$ for the third state.  Next apply one of the following projection operators
\begin{equation}
\begin{array}{lll}
W_{R^1_1}&=&\frac{1}{3}(\ket{e}\bra{e}+\xi\ket{c_+}\bra{c_+}+\xi^{\ast}\ket{c_-}\bra{c_-}),\\
W_{R^2_1}&=&W_{R^1_1}^{\ast}\\
W_{R^4_1}&=&\frac{1}{2}(\ket{e}\bra{e}-\ket{t_0}\bra{t_0}).
\end{array}
\end{equation}
These projections cannot be done unitarily with unit probability but similar to the implementations of $W_{R_2}$ one can use an adaptive protocol to realize them with unit probability.  

\subsection{Anyonic interferometry}
\label{anyinterf}
Let's see how our spin lattice model reproduces these braiding relations given in Sec. \ref{sec:braid}.  First consider braiding of fluxes.  In Sec. \ref{mcharge}, we described how to create a total magnetic charge zero state like Eq. \ref{zmcharge}.  One process we could attempt is to simulate is to create two pairs of total charge zero states, braid a member of one pair around a member of the other, and annihilate.  

Say we begin in the state:
\be
\begin{array}{lll}
&&\ket{\Psi_{\rm in}}=\frac{1}{\sqrt{2}}\sum_{\rho=\pm}\ket{(P^{R_1^+},c_{\rho});(v_{2,4},f_{1,3})}\ket{(P^{R_1^+},c_{-\rho});(v_{2,4},f_{1,4})}\\
&&\frac{1}{\sqrt{3}}\sum_{j=0,1,2}\ket{(P^{R_1^+},t_j);(v_{3,3},f_{2,2})}\ket{(P^{R_1^+},t_j);(v_{3,3},f_{2,4})}
\end{array}
\ee
If we braid the flux $[c]$ at $f_{1,3}$ around the flux $[t]$ at $f_{2,2}$, we obtain the output state
\be
\begin{array}{lll}
&&\ket{\Psi_{\rm out}}=\frac{1}{\sqrt{6}}\sum_{j=0,1,2}\sum_{\rho=\pm}\ket{(P^{R_1^+},c_{-\rho});(v_{2,4},f_{1,3})}\ket{(P^{R_1^+},c_{-\rho});(v_{2,4},f_{1,4})}\\
&&\ket{(P^{R_1^+},t_{j+\rho});(v_{3,3},f_{2,2})}\ket{(P^{R_1^+},t_j);(v_{3,3},f_{2,4})}
\end{array}
\ee
using the relations:  $c_{\rho}t_jc_{-\rho}=t_{j+\rho}$ and $c_{\rho}t_jc_{\rho}t_jc_{-\rho}=c_{\rho}t_jt_{j+\rho}=c_{\rho}c_{\epsilon_{j,j+\rho}}=c_{-\rho}$.  Now we can attempt to fuse the $[c]$ flux anti-flux pair by propagating the flux at $(v_{1,3},f_{1,3})$ one unit to the right.  This is done by preparing the ancilla $f_{1,3}$ in state $\ket{e}$, mapping the flux to $f_{1,3}$ by $\Lambda(v_{2,4},f_{1,3})$ and applying the controlled unitary $Y(f_{1,3},e_{1,4;2,4})$.  If no braiding had been done by the $[t]$ flux then the ancilla would be left in the state $\ket{\phi^+}_{f_{1,3}}$ where $\ket{\phi^{\pm}}=(\ket{c_+}\pm\ket{c_-})/\sqrt{2}$.  What we find is that the ancilla will not be disentangled from the system because the fusion is incomplete.  Rather the ancilla is left in the mixed state $\rho(f_{1,3})=\frac{1}{2}(\ket{c_+}_{f_{1,3}}\bra{c_+}+\ket{c_-}_{f_{1,3}}\bra{c_-})$.  To measure this incomplete fusion, we measure $p=\bra{\phi^-}\rho(f_{1,3})\ket{\phi^-}$.  Before braiding, $p=0$ while afterward $p=1/2$.

Next we examine the braiding of charges.  Braiding of charges around each other acts trivially in the spin lattice quantum double model here because all the operations for creating and moving electric charges are diagonal in the logical basis and hence must commute.  To obtain the other relations first consider the effect of applying a local gauge transformation to a electric charge anti-charge pair separated by a chain of vertices $[v_1,v_2,\ldots v_k]$.  For simplicity we consider that all the edges have the same orientation as the chain, i.e. $e_1=[v_1,v_2], e_2=[v_2,v_3]$ etc., though this is not necessary as in the case of an edge with opposite orientation to the chain one simply takes the inverse group element in the projection.  The action is
\begin{equation}
\fl
\begin{array}{lll}
T_{h}(v_1)&&\ket{(P^R_{\mu,0},e);(v_1,-)}\ket{(P^{R*}_{\nu,0},e);(v_k,-)}\\
&=&\sqrt{\frac{|R|}{|G|}}T_h(v_1)\Big[\sum_{h_1,\ldots h_k \in G}\ket{h_1}_{e_1}\bra{h_1}\otimes \cdots \otimes \ket{h_k}_{e_k}\bra{h_k}   [R(h_1\ldots h_k)^{\ast}]_{\mu,\nu}\Big]T_{h^{-1}}(v_1)\ket{GS}\\
&=&\sqrt{\frac{|R|}{|G|}}\Big[\sum_{h_1,\ldots h_k \in G}\ket{hh_1}_{e_1}\bra{hh_1}\otimes \cdots \otimes \ket{h_k}_{e_k}\bra{h_k}   [R(h_1\ldots h_k)^{\ast}]_{\mu,\nu}\Big]\ket{GS}\\

&=&\sqrt{\frac{|R|}{|G|}}\Big[\sum_{h'_1,\ldots h_k \in G}\ket{h'_1}_{e_1}\bra{h'_1}\otimes \cdots \otimes \ket{h_k}_{e_k}\bra{h_k}   [R(h^{-1}h'_1\ldots h_k)^{\ast}]_{\mu,\nu}\Big]\ket{GS}\\

&=&\sqrt{\frac{|R|}{|G|}}\Big[\sum_{h'_1,\ldots h_k \in G}\ket{h'_1}_{e_1}\bra{h'_1}\otimes \cdots \otimes \ket{h_k}_{e_k}\bra{h_k} \sum_{\kappa}[R(h^{-1})^{\ast}]_{\mu,\kappa}[R(h'_1\ldots h_k))^{\ast}]_{\kappa,\nu}\Big]\ket{GS}\\
&=&\sum_{\kappa}[R(h^{-1})^{\ast}]_{\mu,\kappa}\ket{(P^R_{\kappa,0},e);(v_1,-)}\ket{(P^{R*}_{\nu,0},e);(v_k,-)}\\
&=&\sum_{\kappa}[R(h)]_{\kappa,\mu}\ket{(P^R_{\kappa,0},e);(v_1,-)}\ket{(P^{R*}_{\nu,0},e);(v_k,-)}\\
\end{array}
\end{equation}
where we used the gauge invariance of the ground state.  Similarly we can compute the action of a gauge transformation on the anti-charge, and we find for a general electric charge vacuum state:
\begin{equation}
T_h(v)\ket{M^R;(v,v')}=\ket{R(h)M^R;(v,v')},\quad T_h(v')\ket{M^R;(v,v')}=\ket{M^RR(h^{-1});(v,v')}.
\label{echargemcharge}
\end{equation}
But this local gauge transformation $T_h(v)$ can also be viewed as creating a flux anti-flux pair $h,h^{-1}$ and braiding it in a counterclockwise sence around the electric charge at $v$ followed by annihilation of the flux pair.  The annihilation probability for the flux was zero, meaning the only non trivial action was on the electric charge pair as indicated, hence our model reproduces the correction braiding relations for a flux around a charge. 
In particular, given that we can create the vacuum state for the representation $R$:  $\ket{{\bf 1}_{|R|};(v_{i,j},v_{p,q})}$, we can create the state $\ket{R(h);(v_{i,j},v_{p,q})}$, by applying $T_h(v_{i,j})$, which  is a product of four single qudit unitaries.   How can be measure the effect of this braiding?  
One way to is invert the vertex ancilla steps that prepared the electric charge pair.  For the state $\ket{{\bf 1}_{|R|};(v,v')}$, the inverted sequence will leave the ancilla disentangled from the system, whereas if the initial state is $\ket{R(h);(v,v')}$ then the ancilla will be left in a mixed state.

To actually extract the value of amplitudes for fusion of charges into the vacuum we can use the vertex ancilla prepared in $\ket{\psi_x^+}$ where $\ket{\psi_x^{\pm}}=(\ket{e}_v\pm\ket{h}_v)/\sqrt{2}$, and apply the controlled operation $W(v)$ followed by measurement of the ancilla in the basis $\ket{\psi^{\pm}}$ with outcome $m=\pm 1$.  The outcome distribution satisfies
\[
P(m=1)-P(m=-1)=\Re[\bra{{\bf 1}_{|R|};(v,v')} R(h);(v,v')\rangle]=\frac{\Re[\tr [R(h)]}{|R|}
\]
which is the real part of the fusion amplitude for $R(h)\rightarrow$ vacuum.  Similarly, measuring the ancilla in the basis $\ket{\psi_y^{\pm}}=(\ket{e}_v\pm i\ket{h}_v)/\sqrt{2}$, yields the imaginary part of the fusion amplitude for $R(h)\rightarrow$ vacuum.

\subsection{Elementary operations for quantum computation with anyons}

We now show explicitly how to create anyonic states and perform braiding and fusion operations which are universal for computation.  This section follows closely the work of Mochon \cite{mochon}
who proved two important facts:  first that by working with magnetic charge anyons alone from nonsolvable, non nilpotent groups, universal quantum computation is possible, and second that for some groups that are solvable but not nilpotent, in particular $S_3$, universal quantum computation is also possible if one includes some operations using electric charges.


The first step is to identify the logical basis.  We will encode a qutrit in the three charge $[t]$ magnetic fluxes: $t_0,t_1,t_2$.  We will work with pairs of flux with total trivial flux and will adopt the simplified notation for pure magnetic charges located at positions $r$ and $s$ to $\ket{t_j,t_j^{-1};(r,s)}\equiv  \ket{(P^{R_1^+},t_j);(v,f_i)}\ket{(P^{R_1^+},t_j^{-1});(v,f_j)}$ and for an electric charge pair at positions $r,s$, $\ket{M^R;(r,s)}=\ket{M^R;(v_r,v_s)}$.  Since we will be working with magnetic fluxes that may not have zero charge, the base point vertex $v$ is important but for convenience we suppress it keeping in mind that all fluxes should be valued with respect to some fiducial base point.
The $Z$ basis is $\{\ket{t_j,t_j^{-1}}\}$ and the $X$ basis is $\{\ket{\tilde{t_j},\tilde{t_{j}}^{-1}}\}$ where
\[
\ket{\tilde{t_j},\tilde{t_j}^{-1};(r,s)}=\frac{1}{\sqrt{3}}\sum_{k=0}^2 \xi^{-j} \ket{t_k,t_k^{-1};(r,s)}
\]
The state $\ket{\tilde{t_0},\tilde{t_0}^{-1}}$ is just the vacuum magnetic charge state $\ket{0_{[t]}}$.  The generators of the Pauli group operations on this basis are
\[
X(r,s)=\sum_{j=0}^2 \ket{t_{j+1},t_{j+1}^{-1};(r,s)}\bra{t_j,t_j^{-1};(r,s)},\quad  Z(r,s)=\sum_{j=0}^2 \xi^{j}\ket{t_{j},t_{j}^{-1};(r,s)}\bra{t_j,t_j^{-1};(r,s)}
\]
Encoded information is always stored in flux anti-flux pairs, and the braiding
operations we employ will always braid the pair together thus perserving the
total zero flux condition (see. Eq. \ref{fluxpairbraid}) though there may be a residual electric charge. 
First we describe how to initialize single qudits in the $X$ or $Z$ basis.  The initialization of the state 
$\ket{\tilde{t_j},\tilde{t_j}^{-1}}$ is equivalent to preparation of the vacuum state described in section \ref{mcharge} except where we alter the correction step via the face ancilla $f$ accordingly, i.e. we apply
$[{\bf 1}_{|G|-|[\ell]|}\oplus Z_{|[\ell]|}^{k+j}](f)$ given a measurement outcome $\ket{k_{[\ell]}}_f$. 
To initialize the state  $\ket{t_j,t_j^{-1};(f_1,f_2)}$ we require a means to perform a projection.
The main ingredient necessary is the projection of the qutrits onto the subspace $\mathcal{K}^{t_j^{\perp}}={\rm span}_{\mathbb{C}}\{\ket{t_k,t_k^{-1}},k\neq j\}$.  This can be accomplished with the assistance of an ancilla pair in the vacuum two dimensional electric charge state $\ket{{\bf 1}_{R_2};(r_1,r_2)}$.  Say the initial state of the qutrit is
\[
\ket{\Psi;(1,2)}=\sum_{j=0}^2 c_j \ket{t_j,t_j^{-1};(1,2)}
\]
and we want to project out the $t_0$ component, i.e. we want to project onto $\mathcal{K}^{t_0^{\perp}}$.  To do so we braid flux $1$ around the ancillary electric charge at $r_2$ creating the correlated state
\[
\mathcal{R}^2_{1,r_2}\ket{\Psi}\ket{{\bf 1}_{R_2};(r_1,r_2)}=\sum_{j=0}^2 c_j \ket{t_j,t_j^{-1};(1,2)} \ket{R_2(t_j);(r_1,r_2)}
\]
Next we apply the ancilla assisted local gauge transformation $T_{t^{-1}_0}(r_2)$ so that the joint state is
\[
c_0 \ket{t_0,t_0^{-1};(1,2)} \ket{{\bf 1}_{R_2};(r_1,r_2)}+c_1 \ket{t_1,t_1^{-1};(1,2)} \ket{R_2(c_+);(r_1,r_2)}+c_2 \ket{t_2,t_2^{-1};(1,2)} \ket{R_2(c_-);(r_1,r_2)}
\]
Then we annihilate the electric charge pair and observe whether it fused into the vacuum (using the vertex ancilla protocol described in Sec. \ref{anyinterf}).  The states $\ket{R_2(c_{\pm})}$ always fuse into one dimensional representations (either the vacuum state or the sign representation each with probability $1/2$).  If the charges fuse into the vacuum then we throw the qutrit out, whereas if the outcome is imperfect annihilation then the projection onto the subspace $\mathcal{K}^{t_0^{\perp}}$ is successful.  Similarly, to project onto $\mathcal{K}^{t_j^{\perp}}$ we use the same protocol but with the local gauge transformation $T_{t^{-1}_j}(r_2)$.  By composition of projections we can prepare any basis state beginning in $\ket{0_{[t]}}$, e.g. $\ket{t_1,t_1^{-1};(1,2)}=\mathcal{K}^{t_2^{\perp}}\mathcal{K}^{t_0^{\perp}}\ket{0_{[t]};(1,2)}/\sqrt{||\cdot ||}.$

Let's see how to generate the controlled sum operation $\Sigma((1,2);(3,4))$, where
\[
\Sigma((r,s);(p,q))=\sum_{j=0}^2\ket{t_j,t_j^{-1};(r,s)}\bra{t_j,t_j^{-1};(r,s)}\otimes X^j(p,q)
\]
To realize this we need two primitive braiding operations.
First consider the anicilla assisted operation with the ancilla prepared in $\ket{t_0,t_0^{-1};(5,6)}$,
\be
\begin{array}{lll}
&&\sigma((3,4);(5,6))[\mathcal{R}^2_{5,3}\mathcal{R}^2_{5,4}][\mathcal{R}^2_{3,5} \mathcal{R}^2_{3,6}]\ket{t_k,t_k^{-1};(3,4)}\ket{t_0,t_0^{-1};(5,6)}\\
&=&\sigma((3,4);(5,6))[\mathcal{R}^2_{5,3}\mathcal{R}^2_{5,4}]\ket{t_k,t_k^{-1};(3,4)}X^{2k}\ket{t_0,t_0^{-1};(5,6)}\\
&=&\sigma((3,4);(5,6))\ket{t_{2k}t_jt_{2k}^{-1},t_{2j}t_k^{-1}t_{2k}^{-1};(3,4)}\ket{t_{2k},t_{2k}^{-1};(5,6)}\\
&=&\ket{t_{2k},t_{2k}^{-1};(3,4)}\ket{t_0,t_0^{-1};(5,6)}\\\
\end{array}
\ee
The final swap $\sigma$ on the two pairs is just a classical operation since each pair has total charge zero. 
Second, consider the action of braiding one control flux around the target pair
\be
\mathcal{R}^2_{1,3} \mathcal{R}^2_{1,4}\ket{t_j,t_j^{-1};(1,2)}\ket{t_k,t_k^{-1};(3,4)}=\ket{t_j,t_j^{-1};(1,2)}\ket{t_{2j+2k},t_{2j+2k}^{-1};(3,4)}
\ee
The composition of these two operations realizes the inverse controlled-sum gate:
\be
\begin{array}{lll}
\ket{t_j,t_j^{-1};(1,2)}\ket{t_k,t_k^{-1};(3,4)}&\rightarrow& \ket{t_j,t_j^{-1};(1,2)}\ket{t_{2j+4k},t_{2j+4k}^{-1};(3,4)}\\
&=& \ket{t_j,t_j^{-1};(1,2)}\ket{t_{2j+k},t_{2j+k}^{-1};(3,4)}\\\
&=&[\Sigma((1,2);(3,4))]^{-1} \ket{t_j,t_j^{-1};(1,2)}\ket{t_{k},t_{k}^{-1};(3,4)}\
\end{array}
\ee
Now $[\Sigma((1,2);(3,4))^{-1}]^2=\Sigma((1,2);(3,4))$, hence by two applications we obtain the controlled sum gate.  
Using a controlled-sum gate with the control prepared in state $\ket{t_j,t_j^{-1};(1,2)}$ allows the implementation of an $X^j$ gate on the target.   Imagine that we had an ancillary source prepared in
$\ket{\tilde{t_j},\tilde{t_j}^{-1};(3,4)}$.  Using a controlled-sum gate with the target being this ancilla 
we can perform the one qudit operation $Z^j$ on the control which is verified by considering the action on a complete basis for the control:
\[
\Sigma((1,2);(3,4))\ket{\tilde{t_i},\tilde{t_i}^{-1};(1,2)}\ket{\tilde{t_j},\tilde{t_j}^{-1};(3,4)}=[Z^j
\ket{\tilde{t_i},\tilde{t_i}^{-1};(1,2)}]\ket{\tilde{t_j},\tilde{t_j}^{-1};(3,4)}.
\]
How do we prepare such an ancilla?  We can do this by preparing two ancillary pairs in the state 
$\ket{\Psi}=\ket{0_{[t]};(1,2)}\ket{t_0,t_0^{-1};(3,4)}$ and applying the controlled sum gate
$\Sigma((3,4);(1,2))$.  This prepares a maximally correlated state and if we trace over the pair $(1,2)$, we
obtain the maximally mixed state $\rho_{3,4}=\frac{1}{3}\sum_{j=0}^2 \ket{t_j,t_j^{-1};(3,4)}\bra{t_j,t_j^{-1};(3,4)}=\frac{1}{3}\sum_{j=0}^2 \ket{\tilde{t_j},\tilde{t_j}^{-1};(3,4)}\bra{t_j,t_j^{-1};(3,4)}$.  Now using the methods of \cite{mochon} one can measure $\rho_{1,2}$ in the $X$ basis which will project the state onto 
$\ket{\tilde{t_j},\tilde{t_j}^{-1};(3,4)}$ with probability $1/3$.  This process can be repeated until the desired outcome is obtained.  Such a measurement requires only the preparation of many ancilla pairs prepared in $\ket{0_{[t]}}$ Similary, one can measure in the $Z$ basis with the assistance of many ancilla pairs prepared in $\ket{t_0,t_0^{-1}}$.

Up to now we have shown how to implement the Clifford group operations: controlled-sum, and $X^aZ^b$ as well as preparation in an $X$ or $Z$ eigenstate and measurement in the $X$ and $Z$ basis.
In order to realize universal quantum comptuation we require the ability to perform a non Clifford operation such as the three qutrit Toffoli gate defined by the action:
\[
\begin{array}{lll}
&&T((1,2);(3,4);(5,6))\ket{t_j,t_j^{-1};(1,2)}\ket{t_{k},t_{k}^{-1};(3,4)}\ket{t_{\ell},t_{\ell}^{-1};(5,6)}\\
&=&\ket{t_j,t_j^{-1};(1,2)}\ket{t_k,t_k^{-1};(3,4)}\ket{t_{jk+\ell},t_{jk+\ell}^{-1};(5,6)}
\end{array}
\]
Because the group $S_3$ is solvable we cannot implement the Toffoli gate by braiding magnetic charges alone \cite{Ogburn:98}.  However, as shown by Mochon \cite{mochon}, we can use allowed braiding operations with electric charges to aid in preparing magic states which facilitate the Toffoli gate.  We require two types:
\[
\begin{array}{lll}
\ket{\phi_{M1};(r_1,r_2);(r_3,r_4);(r_5,r_6)}&=&\frac{1}{3}\sum_{j,k=0}^2\ket{t_j,t_j^{-1};(r_1,r_2)}\ket{t_k,t_k^{-1};(r_3,r_4)}\ket{t_{jk},t_{jk}^{-1};(r_5,r_6)}\\
\ket{\phi_{M2};(r_1,r_2);(r_3,r_4)}&=&\frac{1}{3}\sum_{j,k=0}^2\xi^{\delta_{0,j}\delta_{k,0}}\ket{t_j,t_j^{-1};(r_1,r_2)}\ket{t_k,t_k^{-1};(r_3,r_4)}
\end{array}
\]
To realize Toffoli gate on the intial state of three charge pairs
\[
\ket{\Psi;(1,2;3,4;5,6)}=\sum_{a,b,c=0}^2 \beta_{a,b,c}\ket{t_a,t_a^{-1};(1,2)}\ket{t_b,t_b^{-1};(3,4)}\ket{t_c,t_c^{-1};(5,6)}
\]
we append the magic state $\ket{\phi_{M1};(r_1,r_2);(r_3,r_4);(r_5,r_6)}$ and apply
\[
[\Sigma((5,6);(r_5,r_6))][\Sigma((r_3,r_4);(3,4))]^{-1}[\Sigma((r_1,r_2);(1,2))]^{-1}
\]
followed by measurement of $1$ and $3$ in the $Z$ basis with outcomes $m_1,m_3$ and $5$ in the $X$ basis with outcome $m_5$ and subsequent correction gates $X(r_1,r_2)^{m_1}X(r_3,r_4)^{m_3}X(r_5,r_6)^{-m_1m_3}$.  The outcome on the ancilla states is
\[
\sum_{a,b,c=0}^2 \beta_{a,b,c}\xi^{m_5 c}\ket{t_{a},t_{a}^{-1};(r_1,r_2)}\ket{t_{b},t_{b}^{-1};(r_3,r_4)}\ket{t_{ab-m_1b-m_3a+c},t_{ab-m_1b-m_3a+c}^{-1};(r_5,r_6)}
\]
Finally we apply $[Z(r_5,r_6)]^{-m_5}[\Sigma((r_3,r_4);(r_5,r_6))]^{m_1}[\Sigma((r_1,r_2);(r_5,r_6))]^{m_3}$ to realize
\[
\ket{\chi}=\sum_{a,b,c=0}^2 \beta_{a,b,c}\xi^{-m_5 ab}\ket{t_{a},t_{a}^{-1};(r_1,r_2)}\ket{t_{b},t_{b}^{-1};(r_3,r_4)}\ket{t_{ab+c},t_{ab+c}^{-1};(r_5,r_6)}
\]
This set of operations has the action of the Toffoli gate acting on $\ket{\Psi}$ up to a phase $\xi^{-m_5ab}$ on the first two qudits.

The correction gate $C=\sum_{a,b}\xi^{ab}\ket{t_{a},t_{a}^{-1}\ket{t_{b},t_{b}^{-1}}\bra{t_{a},t_{a}^{-1}}\bra{t_{b},t_{b}^{-1}}}$ is in the Clifford group and one might hope that it could be generated from our set of available operations, e.g. using the techniques in \cite{Hostens:04}.  However, to do so requires the single qudit discrete Fourier transform gate which is not accessible using braiding operations.  Hence we use the second magic state $\ket{\phi_{M2};(r_7,r_8;r_9,r_{10})}$ and apply the gates $[\Sigma((r_3,r_4);(r_9,r_{10}))]^{m_1}[\Sigma((r_1,r_2);(r_7,r_8))]^{m_3}$ followed by measurement of $r_7$ and $r_9$ in the $Z$ basis with outcomes $m_7,m_9$.  The outcome is
\[
\ket{\chi}\rightarrow \sum_{a,b,c=0}^2 \beta_{a,b,c}\xi^{-m_5 ab+\delta_{m_7,a}\delta_{m_9,b}}\ket{t_{a},t_{a}^{-1};(r_1,r_2)}\ket{t_{b},t_{b}^{-1};(r_3,r_4)}\ket{t_{ab+c},t_{ab+c}^{-1};(r_5,r_6)}
\]
We can append another magic state $\ket{\phi_{M2};(r_{11},r_{12};r_{13},r_{14})}$ and measure again with outcomes $m_{11},m_{13}$.  Repeating $k$ times the total phase accumulated is $\delta(k,a,b)=\sum_{x=1}^k \delta_{m_{3+4k},a}\delta_{m_{5+4k},b}$.  In order to realize the Toffoli gate we demand that for some $k$, $\delta(k,a,b)=m_5ab\bmod{d} \forall a,b$, where $d$ is the qudit dimension (here $d=3$).  Each step of appending a magic state, applying controlled operations, and measuring applies a phase on the state.
This is akin to picking $k$ balls with $d^2$ ``colors'' which are labeled by the pair $(a,b)$ from an independent and identically distributed probability distribution and placing then into $d^2$ bins each labeled by $(a,b)$.  We find a satisfying distribution if the number of balls in each bin $n_{a,b}=m_5 \bmod{d} \forall a,b$.   The probability that after $k$ trials the condition is not satisfied is (for $d$ prime) approximately $(1-1/d^{d^2})^k$.
The preparation of the magic states follows identically the procedure in \cite{mochon} with frequent use of the projection onto $\mathcal{K}^{t_0^{\perp}}$.

\section{Physical Implementation in atomic spin lattices}

An experimental candidate for realization of the ideas herein is cold trapped atoms in a two dimensional optical lattice (with motion quenched in the third dimension).  Single lattice site occupancy of atoms prepared in motional ground states can be prepared in a region of an optical lattice using a variety of loading techniques \cite{Jaksch}.  In our construction we use an antiferromagnetic array of two type of particles:
system particles or A species, and ancillary particles or B species which will assist in state preparation and generation and dynamical propagation of anyonic excitations.   This bipartite division could be simply a spatial labeling of the same type of physical particle or it could correspond to physically different atomic species or subspaces of the same species. The advantage of using different species for system and ancilla particles is that it could assist in addressability of the controlled unitary operations and measurements in our simulation.  In Fig. \ref{fig:1}b we suggest a suitable 2D lattice architecture for trapping and manipulating the atoms.  By tuning the intensities and phase of the trapping fields, particles can be brought together pairwise with left, right, top, and bottom neighbors to facilitate controlled gate operations.   Single particle coherent control and pairwise entanglement generated by controlled collisional exchange interactions have been experimentally demonstrated with this architecture \cite{Porto:07}.  

\begin{figure}
\begin{center}
\includegraphics[width=\columnwidth]{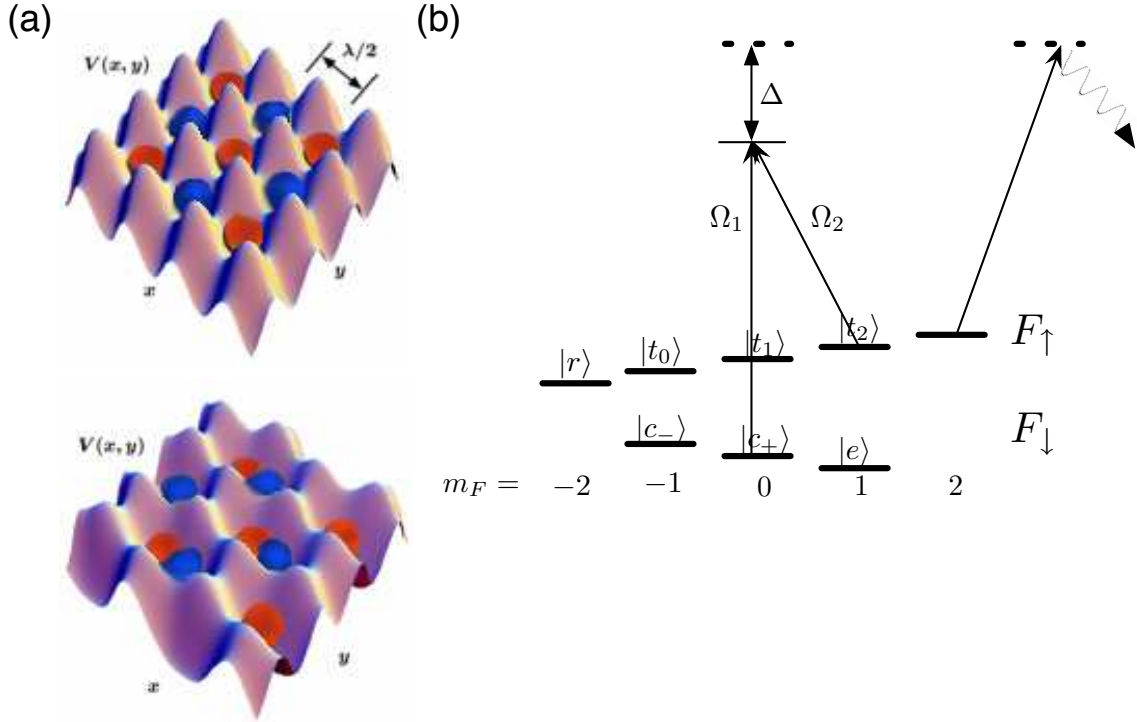}
\caption{ Proposed implementation used atoms trapped in a optical lattice.  (a)  In the 2D retroreflected optical lattice of Ref.  \cite{Porto:06}, when the trapping laser field polarization is in plane, then the lattice has periodicity $\lambda/2$ whereas if the polarization is out of plane, the periodicity is $\lambda$ with a relative shift of $\lambda/2$ resulting in tunable an array of double well potentials.  The upper panel shows the optical lattice for zero field component with out of plane polarization.  In the lower panel, nearest neighbors along one dimension are brought together pairwise by increasing the intensity of the field with out of plane polarization.  Left neighbors or right neighbors are joined depending on the sign of phase of the out of plane polarization light. Top and bottom neighboring pairs can be joined by increasing the out of plane polarization intensity and changing the phase of the field component with polarization in plane.  We note that this architecture allows for global addressing of the A or B species by virtue of distinguishability in the motional states of the two type of lattice wells.  
 (b)  Encoding a $d=6$ qudit into the ground hyperfine levels of a $^{87}$Rb atom.  The logical states are labeled by group elements of $S_3$.  In the figure, the states $\ket{c_+}$ and $\ket{t_2}$ are shown coupled by a Raman laser pulse with Rabi frequencies $\Omega_{1,2}$ by the Hamiltonian $H_{AL}=\Omega_{\rm eff} (e^{i\phi}\ket{c_+}\bra{t_2}+e^{-i\phi}\ket{t_2}\bra{c_+})$, where $\Omega_{\rm eff}=|\Omega_1\Omega_2|/\Delta$ and $\phi=\arg (\Omega_1\Omega_2^{\ast})$.  Arbitrary unitaries in the two dimensional subspace can be generated from this family of Hamitonians, and given a connected coupling graph between basis states, which is provided for here using polarization and frequency selectivity of Raman pulses, any $U\in SU(d)$ can be generated \cite{Brennen:05}.  Projective measurements can be made by mapping population in each sublevel to  $\ket{F_{\uparrow}=2,m_F=2}$ and measuring resonance scattering on the cycling transition $\ket{F_{\uparrow}=2,m_F=2}\rightarrow \ket{F'=3,m_F=3}$ transition.  The state $\ket{c_-}$ can be used for entangling gates via ground state collisions. described by the interaction $H_{\rm int}=U\ket{c_-}_A\bra{c_-}\otimes \ket{c_-}_B\bra{c_-}$ between neighboring atoms A and B. This interaction is diagonal due to total angular momentum and energy conservation.  The state $\ket{r}$ is used as an ancillary state to shelve amplitude in the state $\ket{c_-}$ of atoms that we do not want to participate in a controlled interaction. }
\label{fig:2}
\end{center}
\end{figure}

The six group elements of $S_3$ can be encoded into the ground electronic hyperfine states of a trapped alkali atom with enough available levels.  Candidates include $^{87}$Rb or $^{23}$Na each having $8$ available ground hyperfine states or $^{133}$Cs having $16$ levels.  Arbitrary single qudit operations can be done using Raman laser pulses or microwave pulses that connect pairs of hyperfine states and measurement can be done using a cycling transition (see Fig. \ref{fig:2}).  The atoms are prepared in an optical lattice with spacing $\lambda/2$ where $\lambda$ is an optical wavelength.  At this spacing and for tight lattices, the atoms are well localized and do not interact.  In order to perform controlled unitary operations, one strategy is to bring atoms closer together pairwise and turn on a magnetic field which would generate a trap induced shape resonance on a pair of internal states.  Ideally one would like this collision to be stable in the sense that it would be diagonal: $H_{\rm int}=U\ket{g_j}_A\bra{g_j}\otimes \ket{g_{j'}}_B\bra{g_{j'}}$. This could be provided for by choosing the collision on maximal angular momentum projection states of the ground hyperfine level, chosing e.g. $\ket{g_j}=\ket{g_{j'}}=\ket{F_{\downarrow},m_F=-F_{\downarrow}}$ since a symmetry of the dominate collisional interaction is $m_{F_A}+m_{F_B}$.  Quantum gates using such diagonal collisional interactions generated via trap induced state resonances have been proposed in \cite{Stock:03,Stock:06}.  There it was shown that a robust controlled phase gate between a pair (A,B) of $^{133}$Cs atoms is generated by collisions between the maximally projected angular momentum pair $\ket{F_{\uparrow}, m_F=F_{\uparrow}}_A\ket{F_{\downarrow}, m_F=F_{\downarrow}}_B$ at a particular lattice well spacing.  In order to have high fidelity gates, it is important to work in a parameter regime where there is negligible interactions between other states for two reasons.  First, non maximally projected angular momentum states have elastic collisions which are non diagonal interactions, making a target two qudit unitary difficult to achieve, and second inelastic collisions between states in the upper hyperfine manifold $F_{\uparrow}$ can destroy internal state coherences and convert the internal energy into kinetic energy of the dimer.  

Another possbility is to use the collision gate suggested in \cite{Strauch:07}.  This would work by first using Raman pulses to map the states $\ket{g_j}_{A,B}$ to the first excited vibrational state of lattice wells $A$ and $B$.  Then adjusting the intensity of the trapping fields such that the well minima are tilted with respect to each other and bringing the wells together (but not overlapping) pairwise and finally removing the bias in the wells.  At an appropriate well minima separation, there will be a large tunneling and large collisional shift acquired on atoms in the first vibrational states which are nearly degenerate and near zero tunneling shift for ground vibrational state atoms which are shifted in energy by an amount large compared to the collisional energy.  Then the wells can moved apart again and the population in $\ket{g_j}_{A,B}$ mapped back to the vibrational ground state.  Whatever entangling gate is used, given the ability to generate a phase on one basis state pair, arbitrary two-qudit gates can be generated \cite{Brennen:05}.

In order to perform the quantum simulation here, some amount of addressability of the atoms will be necessary.  This is slightly non trivial as the atoms are separated by $\lambda/2$ and therefore cannot be directly addressed by optical fields.  One solution is to use gradient field spectroscopy wherein a strong local electric field gradient (created e.g. using optical tweezers) is applied in the vicinity of the target atom.  This field will also have some crosstalk with neighbors of the target atom but will shift the energy levels strong enough at the target location such that chirped microwave or Raman operations coupling hyperfine states act trivally at any location other than the target \cite{Zhang:06}. 
To implement two qudit gates on a target pair of atoms A and B we suggest the following strategy:  first for all atoms except A or except B map population in state $\ket{g_j}$ to the spectator internal state $\ket{r}$, second tune the lattice trapping fields to bring A and B together and apply the unitary $U=e^{i\phi \ket{g_j}_A\bra{g_j}\otimes \ket{g_j'}_B\bra{g_j'}}$ in parallel on all nearest neighbor pairs, third invert the mapping to return population to $\ket{g_j}$.  The invertible mapping on non target atoms can be done using STIRAP pulses with shaped intensity profiles as proposed in \cite{Cho:07, Gorshkov:07}.  For example, using a pair of Raman fields that couple states $\ket{c_-}\rightarrow \ket{r}$ (in the encoding of Fig. \ref{fig:2}) one could make the Raman pair out of standing waves of light with a node at the target atom A or B.  For any atom not at a node, there will be an invertible adiabatic mapping  $\ket{c_-}\rightarrow \ket{r}$ which works with fidelity approaching unity.  This technique is particularly well suited to addressing a line of atoms using a 1D standing wave for the STIRAP pulse.  One could also use gradient field spectroscopy to perform the mapping although this may not be as high fidelity due to the cross talk with neighboring atoms.

\section{Conclusions}

We have shown how universal quantum computation by anyonic braiding
can be demonstrated in an optical lattice using the method introduced
in \cite{Aguado}, based on the creation and manipulation of anyons via
controlled interactions of a code lattice with an ancillary species.
In addition, we show how to construct the relevant topological state
using the same method in absence of the suitable Hamiltonian.  The
system considered is the $D ( \mathrm{S}_3 )$ quantum double lattice
model introduced by Kitaev \cite{kitaev_toric}.  In all cases, protocols are thoroughly described and the
experimental techniques required are discussed in detail.

The experimental realization of this scheme relies in methods already
in the literature.  For the procedures described, which exhibit
optimal scaling, the auxiliary species is distributed in a lattice
with the same spacing as the code lattice, so addressability is an
issue, but we propose ways to overcome this problem.
We believe that these techniques are of sufficient interest to warrant
experimental work.  In particular, non-Abelian anyonic statistics can
be demonstrated with our interferometry protocols.

\ack
We acknowledge financial support from the
EU (SCALA) and DFG (MAP and NIM Excellence Clusters).  We thank 
F. Verstraete for comments and discussion.  M.~A.~thanks
S.~Iblisdir, D.~Perez-Garc\'{\i}a, and J.~Pachos for discussions on
the $\mathrm{D}(G)$ models and GKB thanks Ville Lahtinen for reading the manuscript,
suggesting edits, and for several engaging discussions.

\appendix
\section{\quad\quad\quad\quad\quad\quad\quad\quad\quad\quad\quad\quad The quantum double of a finite group}
\label{qdouble}

Given a finite group $G = \{ g_i \}$, the \emph{group algebra}
$\mathbb{C} [G]$ is a complex vector space spanned by basis vectors
$g_i$ (which can be chosen orthonormal to define an inner product.)
Indeed, the multiplication $g_i \otimes g_j \mapsto g_i g_j$ inherited
from $G$ makes $\mathbb{C} [G]$ an algebra.  The unit element $e$ can
be viewed as embedding the complex numbers in $\mathbb{C} [G]$, via
$\lambda \mapsto \lambda e$.
We can also define the algebra $\mathcal{F} (G)$ of complex functions
on $G$, which has a basis given by Kronecker deltas, $P_h : g \mapsto
\delta_{h, \, g}$.  The multiplication in $\mathcal{F} (G)$ can be
recovered from $P_h P_{h'} = \delta_{h, h'} P_h$.  The unit is the map
$1: g \mapsto 1 \forall g$.

Both $\mathbb{C} [G]$ and $\mathcal{F} (G)$ can be endowed with Hopf
algebra structure.  A Hopf algebra $A$ is an algebra where:

(i) Apart from the associative multiplication $m: A \otimes A
\rightarrow A$ and unit $\eta: \mathbb{C} \rightarrow A$ just
described, there exists a comultiplication $\Delta: A \rightarrow A
\otimes A$ and a counit $\varepsilon: A \rightarrow \mathbb{C}$ (a
coalgebra structure) satisfying $( \Delta \otimes \mathrm{id} ) \Delta
= ( \mathrm{id} \otimes \Delta ) \Delta$, and $( \varepsilon \otimes
\mathrm{id} ) \Delta = ( \mathrm{id} \otimes \varepsilon ) \Delta =
\mathrm{id}$, where $\mathrm{id}$ is the identity map on $A$.  For $\mathbb{C} [G]$, these maps are given by
\begin{equation}\label{hopf:comultcg}
 \Delta ( g ) = g \otimes g,
\quad
 \varepsilon ( g ) = 1
 \; ,
\end{equation}
while for $\mathcal{F} (G)$
\begin{equation}\label{hopf:comultfg}
 \Delta ( P_h ) = \sum_{ h_1 h_2 = h } P_{h_1} \otimes P_{h_2},
\quad
 \varepsilon ( P_h ) = \delta_{h, e}
 \; .
\end{equation}
For a general element,
\begin{equation}\label{hopf:comultfg}
 \Delta ( P_h g) = \sum_{ h_1 h_2 = h } P_{h_1}g \otimes P_{h_2}g,
\quad
 \varepsilon ( P_h g) = \delta_{h, e}1
 \; .
\end{equation}
(ii) There exists an antilinear mapping $S: A \rightarrow A$, the
antipode, satisfying $m ( S \otimes \mathrm{id} ) \Delta = m (
\mathrm{id} \otimes S ) \Delta = \eta \varepsilon$.  The antipodes of
$\mathbb{C} [G]$ and $\mathcal{F} (G)$ are determined, respectively,
by $S (g) = g^{-1}$ and $S (P_h ) = P_{ h^{-1} }$.

Moreover, the structures of algebra and coalgebra of $\mathbb{C} [G]$
and $\mathcal{F} (G)$ are dual to each other.  This allows to use
Drinfel'd's quantum double construction to define a quasitriangular
Hopf algebra structure on $\mathcal{F} (G) \times \mathbb{C} [G]$ with
multiplication $P_h g \otimes P_{h'} g' \mapsto \delta_{h, \, g h'
 g^{-1}} P_h g g'$, unit $1 e$, comultiplication $P_h g \mapsto
\sum_{h_1 h_2 = h} P_{h_1} g \otimes P_{h_2} g$,
counit $P_h g \mapsto \delta_{h, e}$, and antipode $P_h g \mapsto
P_{g^{-1} h^{-1} g} g^{-1}$.  This is precisely the quantum double
$\mathrm{D}(G)$ of the finite group $G$.  It is generated as an algebra
by the elements $\{1g|\ g\in G\}$ which form the electric gauge group $G$,
and $\{P_ge|\ g\in G\}$, which are projections onto the set of states 
with flux $g$ in the theory. 

The representation theory of $\mathrm{D}(G)$ is well known.  Each irreducible
representation is determined by a conjugacy class $[ \ell ] = \{ u
\ell u^{-1 }; u \in G \}$ in $G$, and an irreducible representation
$R$ of the centralizer of an element $\ell\in[ \ell ]$, $N_{[ \ell ]} = \{ u \in
G|\ u \ell = \ell u \}$.  Let us label the different elements of the conjugacy class $[\alpha]$
as
\[
[\alpha]=\{h^{[\alpha]}_1,h^{[\alpha]}_2,\ldots,h^{[\alpha]}_k\}
\]
The centralizer $N_{[\alpha]}\subseteq G$, will be defined as the centralizer for the first element $h^{[\alpha]}_1$.
We can relate the different elements of $[\alpha]$ in terms of the first element $h^{[\alpha]}_1$, via a set of representatives of the equivalence classes of $G/N_{[\alpha]}$
\[
X^{[\alpha]}=\{x^{[\alpha]}_{h_1^{[\alpha]}},x^{[\alpha]}_{h_2^{[\alpha]}},\ldots, x^{[\alpha]}_{h_k^{[\alpha]}}|\ h^{[\alpha]}_i =x^{[\alpha]}_{h_i^{[\alpha]}} h^{[\alpha]}_1
(x^{[\alpha]}_{h_i^{[\alpha]}})^{-1}\},
\]
and adopt the convention that $x^{[\alpha]}_{h_1^{[\alpha]}}=e$.
  The carrier space for the irrep
$\Pi^{[\alpha]}_{R(N_{[\alpha]})}$ is
\[
V^{[\alpha]}_{R(N_{[\alpha]})}=\{\ket{h^{[\alpha]}_i,v^{R}_j}|\  h^{[\alpha]}_i\in [\alpha],0\leq j\leq |R|-1 \}.
\]
 The action of the irrep of an element of the Hopf algebra on this space is
\begin{equation}
\Pi^{[\alpha]}_{R(N_{[\alpha]})}(P_hg)\ket{h^{[\alpha]}_i,v^{R}_j}=\delta_{h,g h^{[\alpha]}_i g^{-1}} \sum_{m}\ket{g h^{[\alpha]}_i g^{-1},R(\tilde{g} )_{m,j}\ v^{R}_m}.
\label{algaction}
\end{equation}
The element $\tilde{g}=(x^{[\alpha]}_{gh_i^{[\alpha]}g^{-1}})^{-1}gx^{[\alpha]}_{h_i^{[\alpha]}}$ is constructed from the gauge transformation $g$ and the flux $h^{[\alpha]}_i$ to satisfy $[\tilde{g},h^{[\alpha]}_1]=0$ (verified using the fact that $(x^{[\alpha]}_{gh_i^{[\alpha]}g^{-1}})^{-1}gh_i^{[\alpha]}g^{-1}x^{[\alpha]}_{gh_i^{[\alpha]}g^{-1}}=h_1^{[\alpha]})$ implying that $\tilde{g}\in N_{[\alpha]}$. In this way we see that the the action of the irrep $\Pi^{[\alpha]}_{R(N_{[\alpha]})}(P_h\otimes g)$ on the carrier space is to perform a gauge transformation on the centralizer charge followed by a projection onto the flux conjugated by $g$. 
The dimension of
the carrier space, also known as the quantum dimension of the particle
$\Pi^{[\alpha]}_{R(N_{[\alpha]})}$ is
\[
d^{[\alpha]}_{R(N_{[\alpha]})}=|[\alpha]||R(N_{[\alpha]})|.
\]
The quantum dimension satisfies the sum rule
\[
\sum (d^{[\alpha]}_{R(N_{[\alpha]}})^2=\sum_{[\alpha]}|[\alpha]|^2 \sum_R |R(N_{[\alpha]})|^2=\sum_{\alpha}|[\alpha]|^2 |N_{[\alpha]}|=\sum_{\alpha}|[\alpha]| |G|=|G|^2.
\]
There is some arbitrariness in the choice for representative for
the equivalences classes and the ordering of the elements therein,
however different choices lead to unitarily equivalent representations
of the quantum double.  

The operation of braiding two particles is generated by the monodromy $\mathcal{R}$, which effects a counterclockwise interchange of two particles:
\[
\mathcal{R}=\sigma \circ \Pi^{[\alpha]}_{R(N_{[\alpha]})}\otimes \Pi^{[\alpha']}_{R'(N_{[\alpha']})}(\sum_{h,g\in G}P_g\otimes P_hg).
\]
Here $\sigma$ is the particle interchange operator and the operator in parentheses is the universal $R$-matrix, an element of $\mathrm{D}(G)\times\mathrm{D}(G)$
describing a gauge transformation on the second particle by the flux of the first.  The explicit action on the two particle state space $V^{[\alpha]}_{R(N_{[\alpha]})}\otimes V^{[\alpha']}_{R'(N_{[\alpha']})}$ is:
\[
\mathcal{R}\ket{h^{[\alpha]}_i,v^{R}_j}\ket{h^{[\alpha']}_m,v^{R'}_n}=\sum_{\ell}\ket{h^{[\alpha]}_ih^{[\alpha']}_m(h^{[\alpha]}_i)^{-1},R'(\tilde{h}^{[\alpha]}_i)_{\ell,n}v^{R'}_{\ell}}\ket{h^{[\alpha]}_i,v^{R}_j}
\]
where $\tilde{h}^{[\alpha]}_i=(x^{[\alpha']}_{h_i^{[\alpha]}h_m^{[\alpha']}(h_i^{[\alpha]})^{-1}})^{-1}h_i^{[\alpha]}x^{[\alpha']}_{h_m^{[\alpha']}}$ is defined as above.  Frequently, we are interested in the action of a full counterclockwise braid, $\mathcal{R}^2$, of one particle around another.  For the case of a pure flux braiding another pure flux:
\[
\mathcal{R}^2\ket{h^{[\alpha]}_i,0}\ket{h^{[\alpha']}_m,0}=\ket{(h^{[\alpha]}_ih^{[\alpha']}_m)h^{[\alpha]}_i(h^{[\alpha]}_ih^{[\alpha']}_m)^{-1},0}\ket{h^{[\alpha]}_ih^{[\alpha']}_m(h^{[\alpha]}_i)^{-1},0}
\]
and for a pure flux braiding a pure charge:
\[
\mathcal{R}^2\ket{h^{[\alpha]}_i,0}\ket{e,v^{R}_n}=\sum_{\ell}\ket{h^{[\alpha]}_i,0}
\ket{e,R(h^{[\alpha]}_i)_{\ell,n}v^{R}_{\ell}}.
\]

\section{\quad\quad\quad\quad\quad\quad\quad\quad\quad\quad\quad\quad Some properties of the group $S_3$}
\label{S3rep}

$S_3$ is the group of permutations of three objects, that we label $\{0,1,2\}$, and is the smallest non- Abelian group.  Elements of $\mathrm{S}_3$ are organised in three conjugacy classes,
namely:
\begin{itemize}
\item %
 Identity $e$.
\item %
 Transpositions (reflections) $t_0 = (01)$, $t_1 = (12)$, $t_2 =
 (20)$.
\item %
 3-cycles (rotations) $c_+ = (012)$, $c_- = (021)$.
\end{itemize}

The multiplication rules for $\mathrm{S}_3$ are as follows:
\[
\begin{array}{lll}
 t_i t_i &=& e ,
\qquad
 t_j t_k  = c_{\varepsilon_{j,k}} \quad \mathrm{ for }\ j \neq k ,
\\
 t_i c_\rho &=& t_{i+\rho} ,
\qquad
 c_\rho t_i  = t_{i-\rho} ,
\\
 c_\rho c_\rho   &=& c_{-\rho} ,
\qquad
 c_\sigma c_\tau  = e  \quad \mathrm{ for }\ \sigma \neq \tau .
\end{array}
\label{sthree:eq:multiplication}
\]
together with the trivial operations involving $e$.  Here
$\varepsilon_{j,k} = \pm$ is such that $k = j+ \varepsilon_{j,k}$
(modulo 3, as indicated.)

In particular $t_i^{-1} = t_i$, $c_+^{-1} = c_-$, and conjugation
relations are
\[
\begin{array}{lll}
 t_i t_i t_i &=& t_i ,
\\
 t_j t_k t_j &=& t_i
 \quad
 \mathrm{where\  all\ of }\  i, \, j, \, k \ \mathrm{are\ different,}
\\
\nonumber
 c_\rho t_i c_{-\rho} &=& t_{i+\rho} ,
\\
\nonumber
 t_i c_\rho t_i &=& c_{-\rho}
\\
 c_\rho c_\sigma c_{-\rho} &=& c_\sigma .
\end{array}
\label{sthree:eq:conjugation}
\]

The group has three irreducible representations (irreps): the two one dimensional irreps are the trivial
one $R_1^+(g) = 1$, the signature representation
\begin{equation}
\label{sthree:eq:signature}
 R_1^- ( e )   = +1 = R_1^- ( c_\rho ) ,
\quad
 R_1^- ( t_i ) =   -1 ,
\end{equation}
and the two-dimensional irrep
\[
\nonumber
 R_2 ( e )
=
 \mathbf{1}_2 ,
\quad
 R_2 ( t_k )
=
 \sigma^x
 \exp \left(
        i\, \frac{ 2 \pi }{ 3 } \, k \, \sigma^z
 \right) ,
 R_2 ( c_\rho )
=
 \exp \left(
        i\, \rho\, \frac{ 2 \pi }{ 3 } \, \sigma^z
 \right) ,
\]
Explicitly, 

\[
\begin{array}{lll}
 R_2 ( e )
&=&
 \left(
   \begin{array}{cc}
     1 & 0 \\
     0 & 1
   \end{array}
 \right),
\quad
 R_2 ( c_+ )
=
 \left(
   \begin{array}{cc}
     \xi & 0            \\
     0      &  \xi^{\ast}
   \end{array}
 \right),
\quad
 R_2 ( c_- )
=
 \left(
   \begin{array}{cc}
     \xi^{\ast} & 0      \\
     0            & \xi
   \end{array}
 \right),
\\
 R_2 ( t_0 )
&=&
 \left(
   \begin{array}{cc}
     0 & 1 \\
     1 & 0
   \end{array}
 \right) ,
\quad
 R_2 ( t_1 )
=
 \left(
   \begin{array}{cc}
     0      &  \xi^{\ast} \\
     \xi & 0
   \end{array}
 \right),
\quad
 R_2 ( t_2 )
=
 \left(
   \begin{array}{cc}
     0            & \xi \\
     \xi^{\ast} & 0
   \end{array}
 \right),

\end{array}
\]
where $\xi =e^{ i \, 2 \pi / 3 }$, $\xi^{\ast}=e^{ i \, 4
\pi / 3 }$.

The characters $\chi_R(g)=\tr [R(g)]$ all equal to $\pm 1$ for the one dimensional reps and
\begin{equation}\label{sthree:eq:two_d_irrep_char}
 \chi_{R_2} ( e )   = 2 ,
\quad
 \chi_{R_2} ( t_j ) = 0 ,
\quad
 \chi_{R_2} ( c_\rho ) = -1 .
\end{equation}

The permutation representation for $S_3$ is a set of $6\times 6$
matrices that faithfully represents group left action on the
basis $\{\ket{e},\ket{t_0},\ket{t_1},\ket{t_2},\ket{c_+},\ket{c_-}\}$,
i.e. $L_h\ket{g}=\ket{hg}$.  Similarly, we have unitaries
 for right multiplication
$R_h\ket{g}=\ket{gh}$.  The unitary
matrices satisfy $[L_h,R_{h'}]=0$ and are given by
\[\fl
\begin{array}{lll}
\nonumber
 L_e
&=&
 \left(
   \begin{array}{c|ccc|cc}
     1 &   &   &   &   &   \\ \hline
       & 1 &   &   &   &   \\
       &   & 1 &   &   &   \\
       &   &   & 1 &   &   \\ \hline
       &   &   &   & 1 &   \\
       &   &   &   &   & 1 
   \end{array}
 \right) ,\ 

 L_{ t_0 }
=
 \left(
   \begin{array}{c|ccc|cc}
       & 1 &   &   &   &   \\ \hline
     1 &   &   &   &   &   \\
       &   &   &   & 1 &   \\
       &   &   &   &   & 1 \\ \hline
       &   & 1 &   &   &   \\
       &   &   & 1 &   &  
   \end{array}
 \right) ,\ 

\nonumber
 L_{ t_1 }
=
 \left(
   \begin{array}{c|ccc|cc}
       &   & 1 &   &   &   \\ \hline
       &   &   &   &   & 1 \\
     1 &   &   &   &   &   \\
       &   &   &   & 1 &   \\ \hline
       &   &   & 1 &   &   \\
       & 1 &   &   &   & 
   \end{array}
 \right) ,\ 
\\
 L_{ t_2 }
&=&
 \left(
   \begin{array}{c|ccc|cc}
       &   &   & 1 &   &   \\ \hline
       &   &   &   & 1 &   \\
       &   &   &   &   & 1 \\
     1 &   &   &   &   &   \\ \hline
       & 1 &   &   &   &   \\
       &   & 1 &   &   & 
   \end{array}
 \right) ,\ 

 L_{ c_+ }
=
 \left(
   \begin{array}{c|ccc|cc}
       &   &   &   &   & 1 \\ \hline
       &   & 1 &   &   &   \\
       &   &   & 1 &   &   \\
       & 1 &   &   &   &   \\ \hline
     1 &   &   &   &   &   \\
       &   &   &   & 1 & 
   \end{array}
 \right) ,\ 

 L_{ c_- }
=
 \left(
   \begin{array}{c|ccc|cc}
       &   &   &   & 1 &   \\ \hline
       &   &   & 1 &   &   \\
       & 1 &   &   &   &   \\
       &   & 1 &   &   &   \\ \hline
       &   &   &   &   & 1 \\
     1 &   &   &   &   & 
   \end{array}
 \right),\ 

\end{array}
\]

\[\fl
\begin{array}{lll}
\nonumber
 R_e
&=&
 \left(
   \begin{array}{c|ccc|cc}
     1 &   &   &   &   &   \\ \hline
       & 1 &   &   &   &   \\
       &   & 1 &   &   &   \\
       &   &   & 1 &   &   \\ \hline
       &   &   &   & 1 &   \\
       &   &   &   &   & 1 
   \end{array}
 \right) ,\ 

 R_{ t_0 }
=
 \left(
   \begin{array}{c|ccc|cc}
       & 1 &   &   &   &   \\ \hline
     1 &   &   &   &   &   \\
       &   &   &   &   & 1 \\
       &   &   &   & 1 &   \\ \hline
       &   &   & 1 &   &   \\
       &   & 1 &   &   &  
   \end{array}
 \right) ,\ 

\nonumber
 R_{ t_1 }
=
 \left(
   \begin{array}{c|ccc|cc}
       &   & 1 &   &   &   \\ \hline
       &   &   &   & 1 &   \\
     1 &   &   &   &   &   \\
       &   &   &   &   & 1 \\ \hline
       & 1 &   &   &   &   \\
       &   &   & 1 &   & 
   \end{array}
 \right) ,\ 
\\
 R_{ t_2 }
&=&
 \left(
   \begin{array}{c|ccc|cc}
       &   &   & 1 &   &   \\ \hline
       &   &   &   &   & 1 \\
       &   &   &   & 1 &   \\
     1 &   &   &   &   &   \\ \hline
       &   & 1 &   &   &   \\
       & 1 &   &   &   & 
   \end{array}
 \right) ,\ 

 R_{ c_+ }
=
 \left(
   \begin{array}{c|ccc|cc}
       &   &   &   &   & 1 \\ \hline
       &   &   & 1 &   &   \\
       & 1 &   &   &   &   \\
       &   & 1 &   &   &   \\ \hline
     1 &   &   &   &   &   \\
       &   &   &   & 1 & 
   \end{array}
 \right) ,\ 

 R_{ c_- }
=
 \left(
   \begin{array}{c|ccc|cc}
       &   &   &   & 1 &   \\ \hline
       &   & 1 &   &   &   \\
       &   &   & 1 &   &   \\
       & 1 &   &   &   &   \\ \hline
       &   &   &   &   & 1 \\
     1 &   &   &   &   & 
   \end{array}
 \right). 
\end{array}
\]
Actually, $g \mapsto L_g$ and $g \mapsto R_{g^{-1}}$
 determine the left and right regular representations of $G$,
 respectively.  This is because right multiplication inverts the order of 
 group multiplication $R_gR_{g'}=R_{g'g}$ whereas $\tilde{R}_g=R_{g^{-1}}$
 defines a representation.

Finally, we point out the group $S_3$ has a semi-direct product
structure which may be exploited to simplify physical realizations.
Recall the definition of the semi-direct product. Suppose that we are
given a group $G$ with a normal subgroup $N$, a subgroup $H$, and the
property that any $g\in G$ can be written $g=nh$ for $n\in N$ and
$h\in H$.  Let $\phi$ be the homomorphism $\phi: H\rightarrow {\rm
 Aut}(N)$ where $\phi_h(n)=hnh^{-1}$.  Then $G$ is isomorphic to the
semi-direct product $N\rtimes_{\phi}H$ and the isomorphism identifies
the product $nh\in G$ with the pair $(n,h)\in N\rtimes_{\phi}H$.  We
have $S_3\cong \mathbb{Z}_3\rtimes_{\phi}\mathbb{Z}_2=\langle a,b|
a^3=e,b^2=e,b^q a b^{-q}=a^{2^q}\rangle$.  Here the homomorphism is
specified by $\phi_b(a)=bab^{-1}=a^2$.  Using the notation above we
can choose $\mathbb{Z}_3=\{e,c_+,c_-\}$ and $\mathbb{Z}_2=\{e,t_0\}$,
and any element $g\in S_3$ can be written $g=c_+^r t_0^s$ for $r\in
\{0,1,2\}, s\in \{0,1\}$.  Introducing the basis for group elements
$\{\ket{r}\ket{s}\equiv \ket{c_+^r t_0^s}\}$, i.e. a product basis for
a qutrit and qubit, we have compact representation of the left and
right action operators:
\begin{equation}
\begin{array}{lll}
L_{e}&=&{\bf 1}_3\otimes {\bf 1}_2, \quad  L_{t_0}=F(1,2)\otimes \sigma^x,\quad L_{t_1}=F(0,2)\otimes \sigma^x,\quad L_{t_2}=F(0,1)\otimes \sigma^x,\\
L_{c_+}&=&X^{-1}\otimes {\bf 1}_2, \quad L_{c_-}=X \otimes {\bf 1}_2,\quad R_{e}={\bf 1}_3\otimes {\bf 1}_2, \quad  R_{t_0}={\bf 1}_3\otimes \sigma^x,\\
R_{t_1}&=&X^{-1}\otimes \sigma^-+X\otimes \sigma^+,\quad R_{t_2}=X^{-1}\otimes \sigma^++X\otimes \sigma^-,\\
R_{c_+}&=&X\otimes \ket{0}\bra{0}+X^{-1}\otimes \ket{1}\bra{1},\quad R_{c_-}=X^{-1}\otimes \ket{0}\bra{0}+X\otimes \ket{1}\bra{1}
\end{array}
\end{equation}
where $F(i,j)=(\ket{i}\bra{j}+\ket{j}\bra{i})\oplus 1$ flips two basis
states of the qutrit.  The electric charge creation operators of
Eq. \ref{echargeops} assume a particularly simple form:
$W_{R_1^-}={\bf 1}_3\otimes \sigma^z$ and $W_{R_2}={\rm
 diag}(2,-1,-1)\otimes \ket{0}\bra{0}$, as do the dyonic projectors:
$W_{R_1^1}={\rm
 diag}(\frac{1}{3},\frac{\xi}{3},\frac{\xi^{\ast}}{3})\otimes
\ket{0}\bra{0}$, $W_{R_1^2}=W_{R_1^1}^{\ast}$,
$W_{R_1^4}=\ket{0}\bra{0}\otimes \sigma^z$.

\end{document}